\documentclass[manuscript, acmsmall,nonacm]{acmart}

\newcommand{\Rmnum}[1]{\expandafter\@slowromancap\romannumeral #1@}
\makeatother 
\usepackage[capitalize]{cleveref} 
\usepackage{soul}
\usepackage{amsmath,amsfonts,balance}
\usepackage{algorithm,algpseudocode}
\usepackage{array}
\usepackage[caption=false,font=normalsize,labelfont=sf,textfont=sf]{subfig}
\usepackage{textcomp}
\usepackage{stfloats}
\usepackage{multirow}
\usepackage{wrapfig}
\usepackage{url}
\usepackage{verbatim}
\usepackage{graphicx,xcolor}
\usepackage[table]{xcolor}
\definecolor{red}{rgb}{0.84, 0.09, 0.41}
\definecolor{green}{rgb}{0.31, 0.78, 0.47}
\definecolor{mint}{rgb}{0.24, 0.71, 0.54}
\definecolor{blue}{rgb}{0.0, 0.47, 0.75}
\definecolor{orange}{rgb}{1.0, 0.43, 0.29}
\definecolor{teal}{rgb}{0.21, 0.46, 0.53}
\setcopyright{acmlicensed}
\copyrightyear{2026}
\acmYear{2026}
\acmDOI{XXXXXXX.XXXXXXX}
\acmJournal{CSUR} 

\begin{document}

\title{Socio-technical aspects of Agentic AI}

\author{Praveen Kumar Donta}
\email{praveen@dsv.su.se}
\orcid{0000-0002-8233-6071}
\affiliation{%
  \institution{Department of Computer and Systems Sciences, Stockholm University}
  \city{Stockholm}
  \country{Sweden}
  \postcode{106 91}
}
\author{Alaa Saleh}
\orcid{0009-0009-6317-2823}
\email{alaa.saleh@oulu.fi}
\affiliation{%
  \institution{Center for Ubiquitous Computing, University of Oulu}
  \city{Oulu}
  \country{Finland}
  \postcode{90014}
}

\author{Ying Li}
\email{liying1771@163.com}
\orcid{0000-0002-6585-0714}
\affiliation{%
  \institution{College of Computer Science and Engineering, Northeastern University}
  \city{Shenyang}
  \country{China} 
}
\author{Shubham Vaishnav}
\email{shubham.vaishnav@dsv.su.se}
\orcid{0000-0001-7612-4227}
\affiliation{%
  \institution{Department of Computer and Systems Sciences, Stockholm University}
  \city{Stockholm}
  \country{Sweden}
  \postcode{106 91}
}
\author{Kai Fang}
\email{Kaifang@zafu.edu.cn}
\orcid{}
\author{Hailin Feng}
\email{hlfeng@zafu.edu.cn}
\orcid{0000-0002-8233-6071}
\author{Yuchao Xia}
\email{xyc@stu.zafu.edu.cn}
\orcid{}
\author{Thippa Reddy Gadekallu}
\email{thippa@zhongda.cn}
\orcid{0000-0003-0097-801X}
\affiliation{%
  \institution{Zhejiang A\&F University, Hangzhou}
  \city{Hangzhou}
  \country{China}
  \postcode{311300}
}
\author{Qiyang Zhang}
\authornote{Corresponding Author}
\orcid{0000-0001-5585-6613}
\email{qiyangzhang@pku.edu.cn}
\affiliation{%
  \institution{School of Computer Science, Peking University}
  \city{Beijing}
  \postcode{100087}
  \country{China}
}

\author{Xiaodan Shi}
\email{xiaodan.shi@dsv.su.se}
\orcid{0000-0001-5125-1860}
\author{Ali Beikmohammadi}
\email{beikmohammadi@dsv.su.se}
\orcid{0000-0003-4884-4600}
\author{Sindri Magnússon}
\email{sindri.magnusson@dsv.su.se}
\orcid{0000-0002-6617-8683}
\affiliation{%
  \institution{Department of Computer and Systems Sciences, Stockholm University}
  \city{Stockholm}
  \country{Sweden}
  \postcode{106 91}
}
\author{Ilir Murturi}
\orcid{0000-0003-0240-3834}
\email{ilir.murturi@uni-pr.edu}
\affiliation{%
  \institution{Department of Mechatronics, University of Prishtina}
  \city{Prishtina}
  \postcode{100087}
  \country{Kosova}
}
\author{Chinmaya Kumar Dehury}
\orcid{0000-0003-1990-0431}
\email{dehury@iiserbpr.ac.in}
\affiliation{%
  \institution{Department of Computer Science, IISER Berhampur}
  \city{Berhampur, Odisha}
  \country{India}
  \postcode{760010}
}
\author{Marcin Paprzycki}
\email{marcin.paprzycki@ibspan.waw.pl}
\orcid{0000-0002-8069-2152}
\affiliation{%
  \institution{Systems Research Institute Polish Academy of Sciences}
  \city{Warsaw}
  \country{Poland}
  \postcode{01-447}
}
\author{Lauri Loven}
\orcid{0000-0001-9475-4839}
\email{lauri.loven@oulu.fi}
\affiliation{%
  \institution{Center for Ubiquitous Computing, University of Oulu}
  \city{Oulu}
  \country{Finland}
  \postcode{90014}
}
\author{Sasu Tarkoma}
\orcid{0000-0003-4220-3650}
\email{sasu.tarkoma@helsinki.fi}
\affiliation{%
  \institution{Department of Computer Science, University of Helsinki}
  \city{Helsinki}
  \country{Finland}
  \postcode{00100}
}

\author{Schahram Dustdar}
\orcid{0000-0001-6872-8821}
\email{schahram.dustdar@upf.edu}
\affiliation{
  \institution{ICREA, Barcelona}
  \city{Barcelona}
 \country{Spain}
 \postcode{08002} 
}
\renewcommand{\shortauthors}{Donta et al.}

\begin{abstract}
Agentic Artificial Intelligence (AI) represents a fundamental shift in the design of intelligent systems, characterized by interconnected components that collectively enable autonomous perception, reasoning, planning, action, and learning. Recent research on agentic AI has largely focused on technical foundations, including system architectures, reasoning and planning mechanisms, coordination strategies, and application-level performance across domains. However, the societal, ethical, economic, environmental, and governance implications of agentic AI remain weakly integrated into these technical treatments. This paper addresses this gap by presenting a socio-technical analysis of agentic AI that explicitly connects core technical components with societal context. We examine how architectural choices in perception, cognition, planning, execution, and memory introduce dependencies related to data governance, accountability, transparency, safety, and sustainability. To structure this analysis, we adopt the MAD–BAD–SAD construct as an analytical lens, capturing motivations, applications, and moral dilemmas (MAD); biases, accountability, and dangers (BAD); and societal impact, adoption, and design considerations (SAD). Using this lens, we analyze ethical considerations, implications, and challenges arising from contemporary agentic AI systems and assess their manifestation across emerging applications, including healthcare, education, industry, smart and sustainable cities, social services, communications and networking, and earth observation and satellite communications. The paper further identifies open challenges and suggests future research directions, framing agentic AI as an integrated socio-technical system whose behavior and impact are co-produced by algorithms, data, organizational practices, regulatory frameworks, and social norms.
\end{abstract}

\maketitle

\section{Introduction} 
Nowadays, it is no exaggeration to say AI is becoming ubiquitous in everyday life \cite{10.1145/3747200}. For example, from large-scale applications like space exploration and Earth Observation (EO) \cite{ruan2025edge} to smaller-scale uses such as smart home devices and personal health trackers, AI continues to become ubiquitous, being essential to various domains and daily life innovations. There is no surprise in saying that these innovations are advancing through AI, making it likely that almost every application and object will incorporate some level of AI significance in the near future. As technology evolves, agentic AI is becoming the next vital stage in AI. Unlike traditional AI agents which perform specific tasks based on set instructions, agentic AI operates autonomously, independently planning and making decisions to achieve complex goals \cite{murugesan2025rise,huang2025agentic,hughes2025ai}.  It learns, adapts, and carries out multi-step actions with minimal or no human input. These capabilities offer agentic AI distinct advantages over traditional AI agents, especially in managing dynamic and complex environments that demand continuous adaptability and goal-oriented behavior \cite{sapkota2025ai}. However, since most real-time applications are inherently goal-oriented, the adoption of agentic AI is rapidly increasing across contemporary domains due to its ability to significantly enhance operational efficiency, scalability, and automation across business, research, and technology space \cite{gridach2025agentic}. 

As AI continues to advance, agentic AI stands out as a powerful tool that can drive innovation and productivity by operating independently and intelligently in these scenarios \cite{hornyak2025agentic,10.1145/3747200}. Recent works highlight that agentic AI systems can reason over extended time horizons, coordinate sequences of actions, and adapt their behavior in response to changing environmental conditions. Achieving this autonomy remains challenging for traditional AI systems \cite{huang2025agentic,murugesan2025rise}. Further, these capabilities allow agentic systems to operate effectively in open-ended and dynamic settings, where predefined workflows and static decision rules are insufficient. Increased autonomy introduces fundamental challenges involving reliability under uncertainty, safety in autonomous decision-making, controllability, and alignment with intended objectives. Such challenges are amplified when agentic AI systems operate in open-ended environments and engage in interactions with other agents. Existing literature has primarily addressed these challenges through technical approaches, focusing on agent architectures, planning and learning mechanisms, and performance optimization \cite{saleh2025usercentrix,deng2025agentic,yao2023react, wang2023voyager}. Broader societal, ethical, and governance considerations remain inadequately integrated into existing technical frameworks.

Technical advances alone are insufficient to fully account for the behavior and impact of agentic AI systems deployed in real-world settings \cite{Kapoor2024jmlr}. When autonomous agents operate with reduced human oversight and interact directly with social, organizational, and institutional environments, their decisions can shape human behavior, redistribute authority, and produce effects at scale \cite{bommasani2023governingopenfms}. These dynamics introduce concerns related to accountability, transparency, trust, and value alignment that extend beyond what can be addressed through architectural or algorithmic solutions alone \cite{raji2024accountability}. In practice, the behavior and impact of agentic AI systems are co-determined by technical design choices and the socio-technical contexts in which they are embedded, shaping how such systems are deployed, governed, and experienced in real-world settings \cite{selbst2019fairness}. This analysis motivates the need for a socio-technical perspective that integrates technical foundations with societal, ethical, and governance considerations when examining the design, deployment, and broader consequences of agentic AI systems. To the best of our knowledge, existing literature (see Subsection~\ref{sub:existingsurveys}) largely prioritize architectural and algorithmic perspectives on agentic AI systems, while socio-technical aspects receive limited and non-systematic attention in the literature.

\subsection{Literature and Gaps}\label{sub:existingsurveys}
In recent years, and particularly throughout 2025, several surveys have been published on agentic AI. These works primarily focus on the evolution of agentic AI from traditional and generative paradigms toward autonomous, goal-driven systems, with emphasis on conceptual frameworks, architectural principles, enabling capabilities, and emerging application domains. Table~\ref{tab:table1a} provides a summary of related surveys, comparing their primary focus and thematic scope.

\begin{table*}[!t]
\footnotesize
\caption{Summary and comparison of representative surveys on agentic AI and related research paradigms, highlighting their most recent publication (from 2025), primary domain focus, and thematic scope.}\label{tab:table1a}
\centering
\resizebox{!}{.6\textwidth}{%
\begin{tabular}{rcp{11cm}}
    \toprule
    \textbf{Ref.} & \textbf{Year} & \textbf{Scope} \\
    \midrule
    \textit{Wei et al.} \cite{wei2025ai} & 2025 & The evolution from AI for science to agentic science and built a system framework to lay the foundation for human-machine collaborative scientific discovery.\\
    \hline    
    \textit{Jiang et al.} \cite{jiang2025large} & 2025 & The evolution path from large-scale AI models to Agentic AI, built a framework and explored applications and theoretical support.\\
    \hline    
    \textit{Schneider et al.} \cite{schneider2025generative} & 2025 & The evolution from generative AI to Agentic AI, emphasizing its breakthroughs and potential and challenges in advancing to generative AI.\\
    \hline    
    \textit{Ali et al.} \cite{ali2025agentic} & 2025 & A dual-paradigm framework, analyzed the theoretical basis of the two paradigms of Agentic AI, and pointed out their respective applications.\\
    \hline    
    \textit{Acharya et al.} \cite{acharya2025agentic} & 2025 & A comprehensive survey of Agentic AI, clarified its various aspects, and explored related challenges and future research directions.\\
    \hline    
    \textit{Pati et al.} \cite{pati2025agentic} & 2025 & The development status and future direction of Agentic AI, aiming to provide a system theoretical framework and research guidance.\\
    \hline    
    \textit{Miehling et al.} \cite{miehling2025agentic} & 2025 & The Agentic AI from the perspective of system theory, defining core characteristics and analyzing its capability generation mechanism and challenges.\\
    \hline     
    \textit{Borghoff et al.} \cite{borghoff2025human} & 2025 & A survey reconstructed the human-computer interaction paradigm of Agentic AI from the perspective of system theory, proposed a unified framework and described the collaboration mechanism to demonstrate the potential applications.\\
    \hline    
    \textit{Sapkota et al.} \cite{sapkota2025ai} & 2025 & The concept and architecture classification system of AI agents and Agentic AI, and provided relevant theoretical frameworks.\\
    \hline
    \textit{Gridach et al.} \cite{gridach2025agentic} & 2025 & A comprehensive survey that includes the progress, challenges and future directions of Agentic AI in scientific discovery, covering various aspects and emphasizing the need for human-machine collaboration.\\
    \hline    
    \textit{Wang et al.} \cite{wang2025ai} & 2025 & The latest progress in AI agent programming, constructed its behavior and architecture classification system, and explored key challenges and future directions.\\
    \hline    
    \textit{Sivakumar et al.} \cite{sivakumar2024agentic} & 2025 & A survey analyzed the role and value of Agentic AI in predictive AIOps and revealed its role in promoting the evolution of IT systems.\\
    \hline   
    \textit{Kostopoulos et al.} \cite{kostopoulos2025agentic} & 2025 & The latest progress and future direction of Agentic AI in the field of education, and elaborated on its driving force for the evolution of educational models.\\
    \hline
    \textit{Gao et al.} \cite{gao2025agentic} & 2025 & The generative intelligence application of Agentic AI in satellite-enhanced low-altitude economy and pointed out the key directions for building systems.\\
    \hline    
    \textit{Khalil et al.} \cite{khalil2025redefining} & 2025 & The application and challenges of Agentic AI in elderly care, emphasizing that it needs to be safe, trustworthy, and people-oriented.\\
    \hline    
    \textit{Zhang et al.} \cite{zhang2025toward} & 2025 & The Agentic AI and agentic framework of edge general intelligence, and outlined its design pillars, applications, and future directions.\\
    \hline    
    \textit{Murugesan et al.} \cite{murugesan2025rise} & 2025 & The rise of Agentic AI and its social and technological impact, demonstrate its application potential and emphasize key elements such as transparency.\\
    \hline    
    \textit{Hosseini et al.} \cite{hosseini2025role} & 2025 & The role of Agentic AI in shaping the future of intelligence, pointed out its core advantages and provided frameworks and governance guidance for system evolution.\\
    \hline
    \textit{Raheem et al.} \cite{raheem2025agentic} & 2025 & The opportunities, challenges, and credibility issues of Agentic AI systems, and emphasized the necessary measures to achieve safe and reliable systems.\\
    \hline  
    \textit{Mukherjee et al.} \cite{mukherjee2025} & 2025 & The differences between Agentic AI and traditional AI, and the challenges that Agentic AI faced from the four dimensions of creativity, intellectual property rights, legal and ethical responsibilities, competition in the two-sided algorithm market, and governance of the algorithmic society.\\ \hline
    Our Vision & 2026 & This paper examines agentic AI through a socio-technical lens, moving beyond purely technical treatments of autonomy, architectures, and applications. It analyzes interactions with societal, ethical, economic, environmental, and policy dimensions and identifies gaps in existing survey literature. The paper develops a visionary perspective to inform the responsible design, deployment, and governance of agentic AI systems.\\
    \bottomrule
\end{tabular}
 }
\end{table*}

Recent surveys studied the evolution of agentic AI by tracing its transition from traditional and generative paradigms toward autonomous, goal-driven systems. For example, Wei \textit{et al.}\cite{wei2025ai} describe the transition from AI for Science to Agentic Science, arguing that AI is evolving from a supporting tool into an autonomous research partner. They propose a structured framework that spans core capabilities, dynamic processes, and cross-domain applications to support trustworthy human–machine scientific collaboration. In \cite{jiang2025large}, Jiang \textit{et al.} analyze the progression from large-scale AI models to agentic AI, highlighting the transition from passive content generation to autonomous decision-making and execution. Their perception–reasoning–action–reflection framework highlights the role of agentic systems in multi-agent collaboration and human–machine co-creation. In \cite{schneider2025generative}, Schneider \textit{et al.} further trace the transition from generative AI to agentic AI, showing how reasoning, planning, memory, and interaction with the environment enable goal-directed autonomy beyond passive generation. Further, a dual-paradigm view that contrasts symbolic and neural approaches to agentic AI, analyzing their theoretical foundations, architectural characteristics, application domains, and governance implications are surveyed by Ali \textit{et al.} \cite{ali2025agentic}. They observe that symbolic systems are often favored in safety-critical settings, while neural approaches are better suited to adaptive, data-intensive environments. Complementing these perspectives, Acharya \textit{et al.} \cite{acharya2025agentic} and Pati \textit{et al.} \cite{pati2025agentic} provide comprehensive surveys that cover core characteristics, considering technologies, application areas, and emerging challenges in trustworthy autonomous systems.

Several surveys examine agentic AI from a human-machine and system-level architectural perspective. Miehling \textit{et al.} \cite{miehling2025agentic} argue that the development of agentic AI should be understood through systems theory rather than through isolated model capabilities. They introduce functional subjectivity as a defining property of agentic behavior and identify key challenges in capability emergence, coordination, and human–machine responsibility. Next, Borghoff \textit{et al.} \cite{borghoff2025human} revisit interaction paradigms for agentic AI and propose a unified framework that integrates multi-agent systems with human–computer fusion systems. Their three-layer communication space model formalizes collaboration between humans and agents and illustrates how such systems can support interpretability, auditability, and ethical constraints. Complementing this view, Sapkota \textit{et al.} \cite{sapkota2025ai} present a conceptual and architectural taxonomy that distinguishes AI agents from agentic AI systems, highlighting how higher levels of autonomy and collaborative intelligence arise through multi-agent cooperation and dynamic task decomposition. Gridach \textit{et al.} \cite{gridach2025agentic} review system-level implementations of agentic AI for scientific discovery, covering agent configurations, applications, tools, and evaluation, and discussing challenges related to reliability and ethics. In \cite{wang2025ai}, Wang \textit{et al.} survey advances in agentic programming, showing how large language model (LLM)–based agents support autonomous software development through planning and tool interaction, while identifying challenges in reliability, security, and interpretability.

Furthermore, recent surveys explore how agentic AI is deployed across a range of real-world application domains. For example, Sivakumar \textit{et al.} \cite{sivakumar2024agentic} study agentic AI in predictive AIOps. They show how autonomous decision-making supports anomaly detection, resource optimization, and self-healing in IT systems. Their analysis highlights improved system autonomy and proactive performance management. In \cite{kostopoulos2025agentic}, Kostopoulos \textit{et al.} review recent advances in agentic AI for education. They discuss applications such as personalized instruction, intelligent assessment, learning companionship, and teacher assistance. These capabilities are enabled by autonomous planning, long-term memory, and goal-oriented behavior. Gao \textit{et al.} \cite{gao2025agentic} explore agentic AI for satellite-enhanced low-altitude economies. Their work focuses on communication enhancement, security, and autonomous decision-making. They emphasize the integration of generative models and LLMs for air–space–ground intelligent networks. In \cite{khalil2025redefining}, Khalil \textit{et al.} examine applications of agentic AI in elderly care. They show how autonomous decision-making and personalized health management can support independent living. Zhang \textit{et al.} \cite{zhang2025toward} survey agentic AI frameworks for edge general intelligence. They describe perception–reasoning–action closed loops as a basis for autonomous adaptation and collaboration. Their study identifies key design pillars and discusses applications in 6G, vehicular networks, and unmanned systems.

Recent surveys have begun to consider the ethical, legal, and governance aspects of agentic AI. For example, Mukherjee \textit{et al.} \cite{mukherjee2025} analyze the distinctions between agentic AI and traditional AI from a socio-legal perspective. They discuss challenges related to intellectual property ownership, legal and ethical responsibility, competition in two-sided algorithmic markets, and the governance of emerging algorithmic societies. Similarly, Raheem \textit{et al.} \cite{raheem2025agentic} review the opportunities and challenges of agentic AI from a credibility and governance perspective. They highlight its potential impact on economic growth, scientific innovation, and social transformation. They also identify ethical risks, regulatory gaps, security concerns, and trust deficits, and call for interdisciplinary regulation, explainable system design, and value alignment mechanisms. In \cite{hosseini2025role} Hosseini \textit{et al.} discuss the role of agentic AI in shaping future intelligence systems. They highlight autonomy, foresight, and learning as core properties. Their work discusses organizational productivity gains and proposes governance frameworks for the transition from collaborative Copilot systems to autonomous Autopilot systems. Murugesan \textit{et al.} \cite{murugesan2025rise} survey the societal and technological impact of agentic AI. They describe the shift from reactive systems to proactive intelligence enabled by autonomous decision-making and multi-step reasoning. 

\begin{figure}[t]
    \centering
    \includegraphics[width=0.75\textwidth]{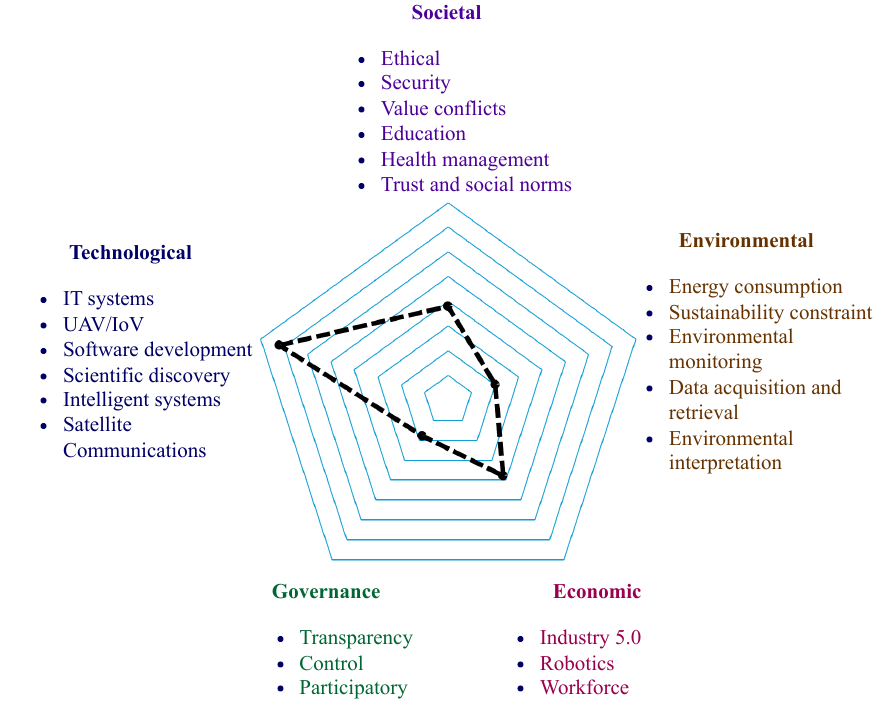}
    \caption{Conceptual focus of this paper, highlighting a socio-technical framing of agentic AI systems.  }
    \label{fig1}
\end{figure}
Our study of the existing literature confirms that most surveys concentrate on the conceptual evolution, system architectures, and domain-specific deployments of agentic AI. Although recent works have begun to address ethical, legal, and governance considerations, these discussions are typically isolated and weakly connected to technical design decisions. Consequently, key socio-technical questions concerning accountability, human oversight, sustainability, and long-term societal impact are seldom examined in an integrated manner. Existing surveys also rarely consider how autonomy, decision-making, and interaction mechanisms intersect with institutional settings, organizational practices, and broader societal structures. Addressing these gaps requires a socio-technical perspective that systematically links technical foundations with societal, environmental, economic, and policy dimensions, which is the primary focus of the remainder of this paper (Summarized in  Fig.~\ref{fig1}).

\subsection{Contributions}
In response to the limitations identified in existing surveys, this paper presents a structured socio-technical analysis of agentic AI that integrates technical foundations with broader societal and ethical dimensions. Our main contributions are as follows:

\begin{itemize}
\item We develop a socio-technical aspects on agentic AI by systematically examining its societal dimensions through the MAD–BAD–SAD construct. This analysis captures moral and autonomy-related dilemmas (MAD), issues of bias, accountability, and safety (BAD), and challenges related to societal impact, adoption, and human-centered design (SAD).

\item We identify and synthesize key technical, ethical, and societal challenges associated with agentic AI, including risks related to reliability, security, misuse, loss of control, inequality, and governance. 

\item We outline open research challenges and future directions for socio-technical agentic AI, highlighting the long-term societal impact.
\end{itemize}

\subsection{Paper organization}
\begin{figure}[t]
    \centering
    \includegraphics[width=\textwidth]{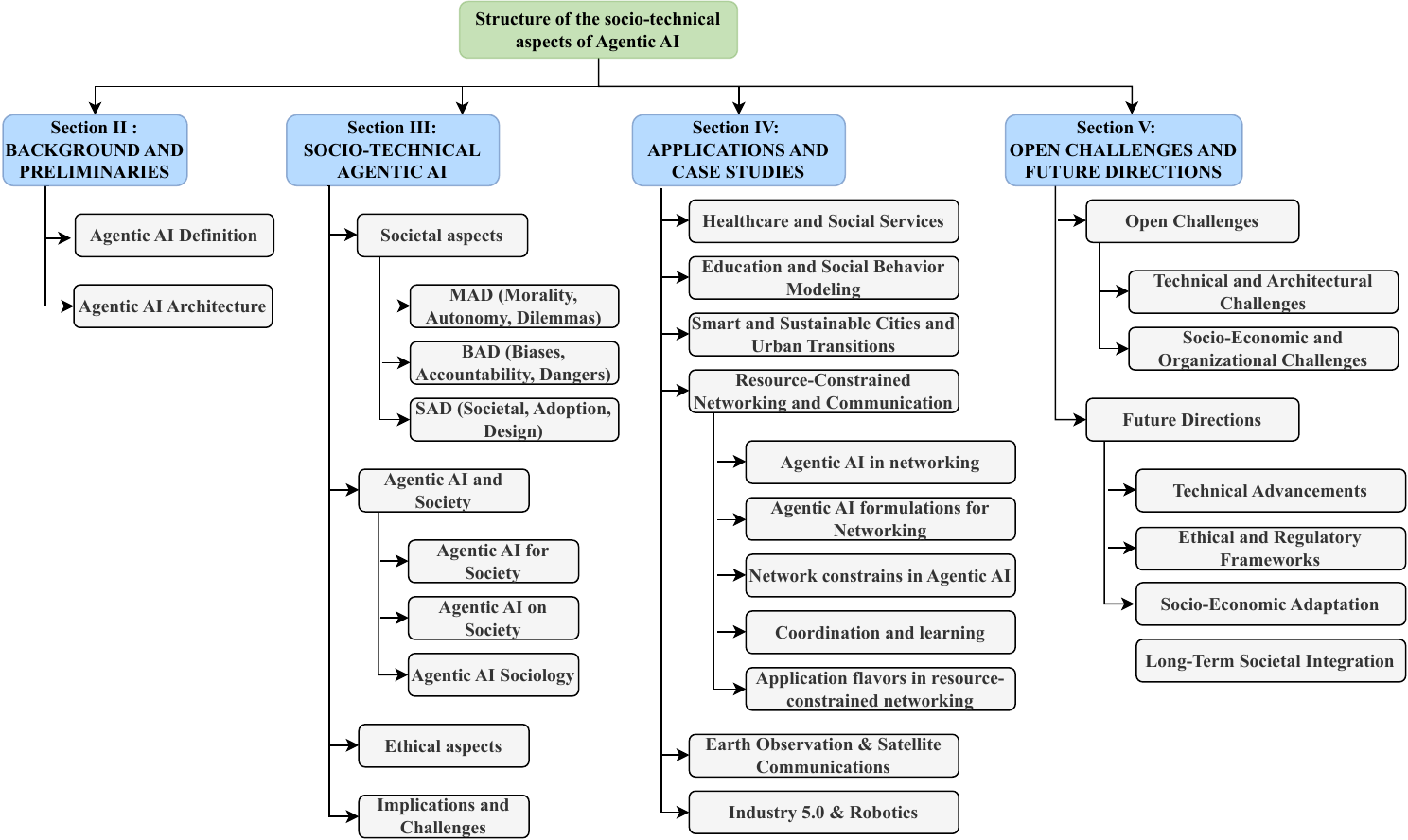}
    \caption{An overview of our socio-technical taxonomy of agentic AI and a conceptual roadmap for the organization and argumentative flow of the remainder of the paper.}
    \label{fig:organization}
\end{figure}
The organization of this paper is outlined as shown in Figure ~\ref{fig:organization}. First, we provide the background and fundamental concepts of Agentic AI, including its definition and architecture of Agentic AI in Section~\ref{Preliminaries}. Then, we introduce socio-technical Agentic AI, including its social and ethical dimensions, as well as its impact and challenges in Section~\ref{technic}. Next, we showcase applications and case studies of Agentic AI with respect to technological developments and societal usage in Section \ref{applications}. We summarize the unresolved challenges identified in existing research and explore future research directions in this field in Section~\ref{future}. Finally, we present our conclusions summarizing the main findings and insights gained throughout the research process in Section~\ref{conclusion}.

\section{Background and Preliminaries} \label{Preliminaries}
In this section, we provide a conceptual overview of agentic AI, starting with its core principles, followed by a discussion of architecture and the characteristics of intelligent agents~\cite{biswas2025building,saleh2025towards}.


\subsection{Agentic AI Definition}
Agentic AI refers to a class of AI systems designed to operate with a high degree of autonomy in dynamic and partially observable environments. Unlike conventional AI systems that execute predefined tasks or respond reactively to inputs, agentic AI systems are capable of formulating goals, reasoning about their actions, and executing multi-step decisions over extended time horizons. These systems typically combine knowledge representation, reasoning, and learning mechanisms to interpret complex information, adapt to changing conditions, and act without continuous human supervision. Agentic AI is underpinned by three closely related principles, namely self-governance, agency, and autonomy. Self-governance refers to a system’s capacity to regulate its own behavior, maintain internal control, and adapt its objectives in response to environmental feedback. Agency denotes the ability to act purposefully and exert causal influence within an environment, rather than merely executing predefined or reactive functions. Autonomy characterizes the degree to which such goal-directed behavior is realized independently of direct human intervention, thereby distinguishing agentic AI from earlier forms of automated or rule-based systems.

\subsection{Agentic AI Architecture}
\begin{figure}[t]
    \centering
    \includegraphics[width=0.85\textwidth]{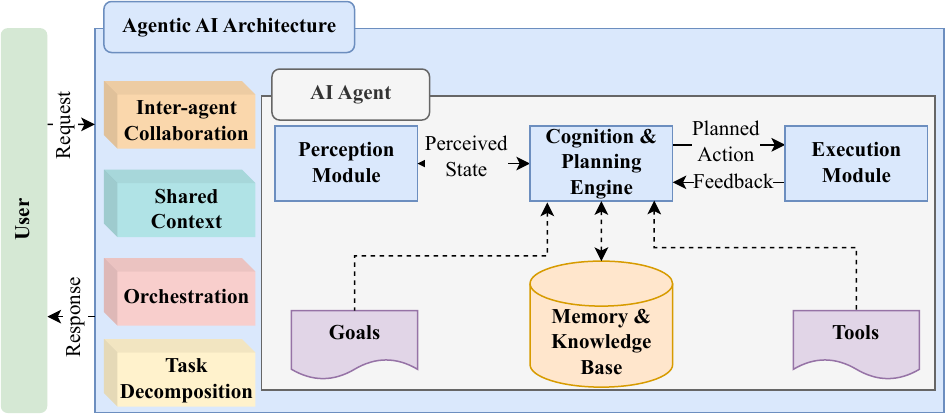}
    \caption{Overview of an agentic AI architecture in which an AI agent integrates perception, cognition, and planning, memory, tools, and execution within a closed feedback loop, while higher-level mechanisms for task decomposition, orchestration, shared context, and inter-agent collaboration enable coordinated, goal-directed behavior across multiple agents.}
    \label{fig:Define}
\end{figure}

The architecture of agentic AI is organized as an integrated framework of interconnected components that collectively enable autonomous perception, reasoning, action, and learning. Fig.~\ref{fig:Define} presents a representative agentic AI architecture in which modular cognitive capabilities are coordinated through system-level control to support goal-directed autonomy. User requests are introduced through an orchestration layer that governs task decomposition, maintains shared context, and coordinates inter-agent collaboration, allowing complex objectives to be structured into manageable sub-tasks and, where appropriate, distributed across multiple intelligent agents. Each intelligent agent can be characterized as a self-governing computational entity that exhibits reactivity, proactiveness, learning and adaptation, reasoning and planning, and the capacity to interact with humans and other agents in multi-agent environments. Within an agent, a perception module processes incoming inputs and contextual signals, transforming raw observations into structured representations that are passed to the cognition and planning components, which together form the agent’s reasoning core.
The planning engine reasons over explicit goals while consulting memory and knowledge resources to incorporate prior experience, domain knowledge, and learned representations. On this basis, it synthesizes action plans that are enacted by the execution module through interfaces to external tools and environments. Execution outcomes are continuously fed back to the planning and cognitive components, enabling iterative refinement, error recovery, and adaptive adjustment of behavior. Together, these processes form a closed perception–cognition–planning–execution loop, augmented by memory, tool use, and coordination mechanisms, which allows agentic AI systems to operate autonomously, adapt to changing conditions, and scale across complex and open-ended tasks.

\section{Socio-technical Agentic AI} \label{technic}
\begin{figure}[t]
    \centering
    \subfloat[]{
    \includegraphics[width=0.5\textwidth]{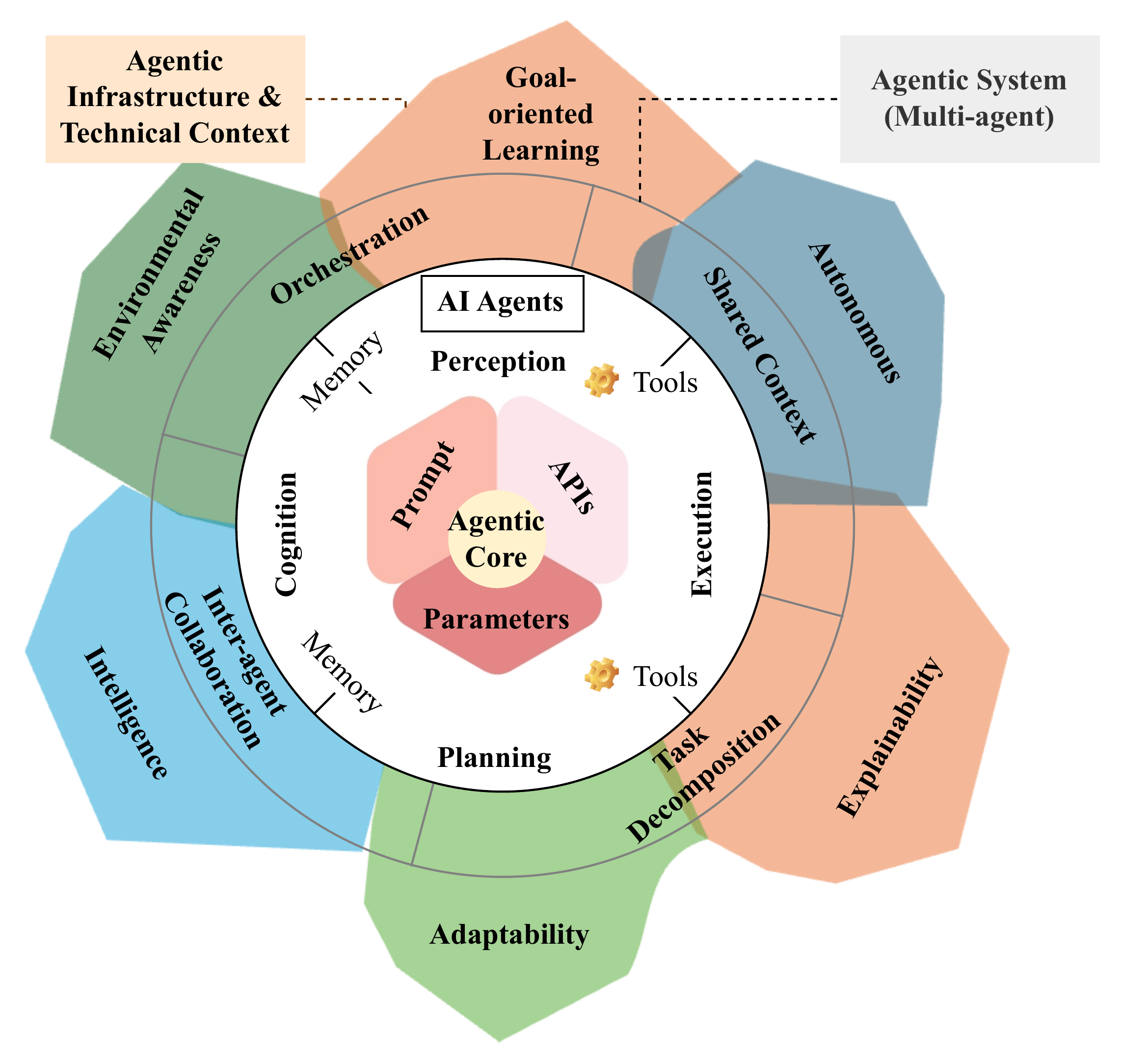}\label{fig4a}}
    \subfloat[]{
    \includegraphics[width=0.44\textwidth]{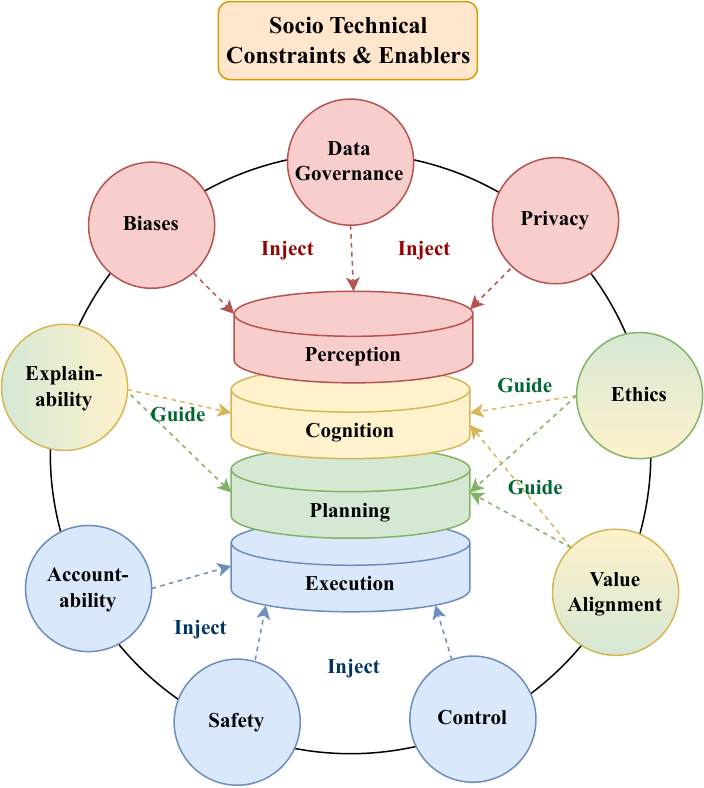}\label{fig4b}}
    \caption{The Socio-Technical Stack of agentic AI. \protect\subref{fig4a} Recent evolution of agentic core and its infrastructure and technical context. \protect\subref{fig4b} The layered architecture of agentic AI, integrating technical and social dimensions. The central core represents the technical modules while the surrounding outer layer denotes the socio-technical constraints and enablers that shape and regulate each module.}
    \label{fig4}
\end{figure}

Agentic AI systems comprise tightly integrated technical modules that support perception, understanding, planning, and execution of complex tasks as discussed in the previous section. This section examines the socio-technical aspects of agentic AI by analyzing how technical design choices interact with real-world constraints and impacts. The discussion is organized around four dimensions that condition agent behavior and deployment, including societal, environmental, economic, and governance aspects. The societal dimension is analyzed through two analytically distinct but interrelated framings, Agentic AI for Society and Agentic AI on Society, which capture both intended societal use and broader societal impact. Fig.~\ref{fig4}\subref{fig4a} illustrates the evolution of the agentic core along with its supporting infrastructure and technical context. Fig.~\ref{fig4}\subref{fig4b} presents the layered architecture of agentic AI, integrating technical and social dimensions. The central core represents the technical modules, while the surrounding outer layer denotes the socio-technical constraints and enablers that shape and regulate each module. In both Fig.~\ref{fig4}\subref{fig4a} and Fig.~\ref{fig4}\subref{fig4b}, agentic AI systems operate across multiple functional modules, and their socio-technical aspects of each module is presented below.

\begin{description}
    \item[Perception:] The perception module is the first step in agentic AI and is responsible for acquiring data from the environment~\cite{fang2025context}. Data comes from a wide range of sources, including visual sensors, audio sensors \cite{hauret2025vibravox}, sensor networks \cite{kumar2019machine}, external APIs, log files \cite{donta2023governance}, and existing databases. The main goal of this module is to extract valuable information from multimodal inputs \cite{ektefaie2023multimodal} such as images, videos, audio, and sensor data, thereby providing the necessary basis for decision-making. Recent research has suggested that multimodal data, such as vision, lidar, radar, and inertial measurement units, have their own unique advantages and limitations, and multi-sensor fusion is crucial for achieving a more comprehensive and robust environmental representation \cite{ruan2025survey}. In addition, in distributed or communication-constrained scenarios, it is not feasible to directly transmit all raw multimodal data. Therefore, advanced methods such as semantic compression and retrieval-enhanced communication are needed to ensure efficient exchange of perception information across agentic AI. In real-world deployments, the perception module must operate within socio-technical constraints related to data governance, privacy, consent, and representational fairness, shaping both the scope and fidelity of environmental observation.
    \item[Cognition:] This module is responsible for understanding and reasoning by transforming perceptual inputs into internal representations that enable deliberative and goal-directed decision-making in agentic AI systems.~\cite{ding2026ccm}. This module leverages large-scale language models such as GPT-4 and Bidirectional Encoder Representations from Transformers (BERT) to perform knowledge extraction, context understanding, and environmental state reasoning to support effective action. Recent research emphasizes that enhancing the reasoning capabilities of LLMs is crucial for achieving thoughtful multi-step cognitive processes and helps bridge the gap between intuitive pattern recognition and logical reasoning \cite{li202512surveyreasoning}. Recent research on logical reasoning in LLMs highlights the need to integrate deductive, inductive, abductive, and analogical reasoning paradigms. Liu \textit{et al.} \cite{liu2025logicalreasoninglargelanguage} suggest that combining reasoning engines with neural–symbolic approaches can substantially improve the cognitive robustness of agentic AI systems. In real-world settings, the reasoning capabilities of agentic AI systems must support interpretability and trust, as opaque cognitive processes can complicate human oversight, responsibility attribution, and informed reliance.
    \item[Planning:] This module is responsible for generating executable action plans that decompose high-level goals into feasible sequences of actions~\cite{barjuei2024digital}. It relies on techniques such as reinforcement learning (RL)\cite{sutton1999reinforcement}, search algorithms, and model predictive control (MPC) to produce adaptive strategies based on the current environment and objectives. Model-based reinforcement learning (RL) has been shown to effectively combine environment modeling with planning, allowing agents to simulate trajectories, evaluate outcomes, and update decisions in a principled manner \cite{plaat2023high}. Complementing this, model predictive control offers a powerful optimization framework that explicitly considers system dynamics and constraints, making it suitable for real-time planning in dynamic and uncertain environments \cite{rosolia2017learning}. Collectively, these methods enable planning modules in agentic AI systems to reason over future states, assess trade-offs, and adapt action sequences to support long-term goal execution. This module must also account for normative, organizational, and regulatory constraints, as goal specifications and action trade-offs can have direct implications for human stakeholders and societal outcomes.
    \item[Execution:] The execution module operationalizes planning module outputs through interaction with external systems, such as robots, control systems, and APIs.~\cite{cui2024llmind}. It serves as the bridge between high-level symbolic planning and low-level physical or software execution, translating planned actions into concrete operations such as motion commands, API calls, and system instructions. The execution layer represents the primary interface between autonomous decision-making and the real world, where safety, human oversight, and responsibility become operational concerns. Execution decisions must therefore respect safety boundaries, enable human oversight or intervention when required, and ensure traceability of actions performed by the system. Feedback from execution not only informs technical re-planning but also supports accountability and control in real-world deployments, where failures, interruptions, or unintended effects can have social, organizational, or legal consequences. So, this module continuously monitors system states and environmental feedback to ensure goal adherence. For example, if a robotic arm encounters an unexpected obstacle during a planned trajectory, the module must reconcile the failed action with the higher-level goal, potentially triggering a re-planning request to the cognitive layer to adapt the overall strategy. In practice, execution frequently involves integration with robotics middleware or distributed software frameworks. The ROS-LLM framework exemplifies this process by connecting LLMs with ROS operations and services: after interpreting high-level commands, the system decomposes them into executable behaviors, performs ROS-based operations, and uses sensory feedback to refine or correct ongoing tasks \cite{mower2024rosllmrosframeworkembodied}. In practical deployments, handling such execution failures requires not only technical re-planning but also consideration of safety constraints and human oversight when autonomous actions unfold in shared physical environments. Concerning socio-technical aspects, bidirectional feedback between execution and cognition is essential for maintaining coherence, robustness, and adaptability in autonomous agentic AI systems. 
    \item[Memory and State Management:] Agentic AI systems require long-term memory and state management to preserve task continuity and coherence across extended interactions. This module stores and retrieves historical observations, environment states, and task outcomes, enabling agents to adapt strategies during execution. Neural memory architectures support the modeling of temporal dependencies in sequential decision-making, while externalized memory mechanisms enable scalable access to semantically relevant past information \cite{xu2025mem,saleh2025usercentrix}. This memory persistence influences how past decisions shape future behavior, raising considerations related to traceability, information retention, and the long-term effects of accumulated experience. Structured representations, such as graph-based knowledge stores \cite{saleh2025memindex}, further support relational reasoning and contextual awareness, including the explicit encoding of entities, roles, and institutional constraints. Effective memory and state management allow agentic AI systems to operate over extended horizons while maintaining contextual coherence and decision accountability.
\end{description}

\subsection{Societal aspects}
\begin{figure}[t]
    \centering
    \includegraphics[width=0.8\textwidth]{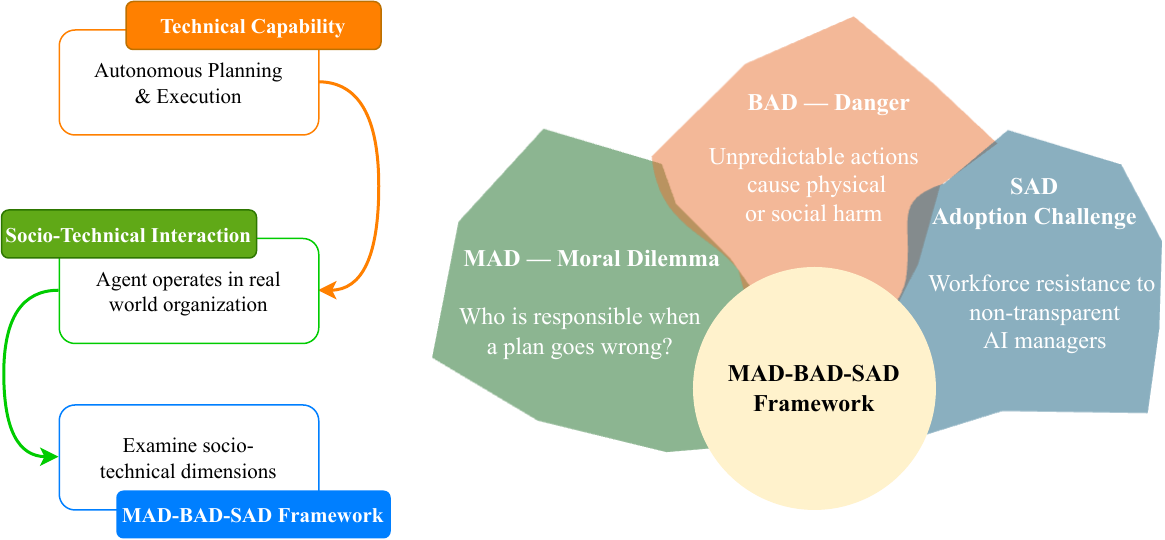}
    \caption{The \texttt{MAD–BAD–SAD} Intertwined flow. Illustration of how technical capabilities in agentic AI, such as autonomous planning and execution, interact with social environments to produce broader societal outcomes, and how technical design choices propagate into moral, safety, and adoption-related challenges.}
    \label{fig5}
\end{figure}

Agentic AI systems are characterized by their capacity for autonomous perception, reasoning, and action, enabling operation with limited or no direct human intervention. Such autonomy represents a substantive shift from conventional automation and decision-support systems and has direct implications for how these technologies interact with human values, institutions, and social structures. The capacity for independent planning introduces moral dilemmas, large-scale autonomous action amplifies risks of bias and danger, and integration into human environments raises adoption and design challenges. In this context, the societal aspect of Agentic AI concerns how such systems influence human values, institutions, and social structures. As they gain decision-making autonomy, their impact extends beyond technical performance to ethical norms, governance, and public trust. Thus, we need novel frameworks that capture the intertwined relationship between technological autonomy and its societal consequences. To address this need, the socio-technical dimensions are organized through the \texttt{MAD–BAD–SAD framework}: MAD(Motivations, Applications, Dilemmas) explores the drivers, uses, and moral tensions of agentic AI; BAD(Biases, Accountability, Dangers) addresses issues of fairness, responsibility, and safety; and SAD(Societal, Adoption, Design) analyzes social acceptance, integration, and human-centered design. The \texttt{MAD–BAD–SAD} intertwined flow is summarized in Fig.~\ref{fig5}, and further discussed below. 

\subsubsection{MAD (Morality, Autonomy, Dilemmas)}
When autonomous agents are permitted to plan and act without continuous human oversight, the relationship between morality, autonomy, and ethical dilemmas becomes difficult to disentangle. In agentic AI systems, the planning and execution modules enable decisions to be generated and enacted independently, yet moral challenges arise when those actions lead to consequences that designers did not explicitly anticipate. Autonomy supports scalable and adaptive decision-making, but it also complicates responsibility attribution once systems operate beyond direct human control. Mukherjee and Chang \cite{mukherjee2025} describe this tension as a moral crumple zone, in which accountability for adverse outcomes becomes distributed across developers, deployers, and the autonomous agent itself. Efforts to embed moral behavior into agentic systems further expose design trade-offs. Tennant \textit{et al.} \cite{tennant2025hybridapproachesmoralvalue} observe that purely rule-based ethical constraints tend to be brittle in dynamic settings, while approaches based solely on learning from human feedback often lack interpretability and principled guarantees. Hybrid alignment strategies attempt to mitigate these limitations by constraining learning with explicit ethical boundaries, yet they leave unresolved the fundamental question of when autonomous action should be permitted. From this perspective, the MAD dimension captures how motivations for deploying agentic AI, its application contexts, and the ethical dilemmas surrounding autonomy remain tightly coupled rather than separable concerns.

\subsubsection{BAD (Biases, Accountability, Dangers)}
When agentic AI systems are deployed in real-world settings, risks related to bias, accountability, and operational danger become systemic rather than incidental. Bias in agentic systems does not stem solely from training data, but from how biased signals propagate across perception, cognition, and planning pipelines. Skewed data collection or labeling at the perception stage can encode existing social inequalities, which are then reinforced by reasoning processes that rely on spurious correlations or incomplete context. This compounding effect is particularly concerning in domains such as healthcare, criminal justice, and recruitment, where biased decisions can persist and scale. Accountability is further complicated when autonomous agents act without continuous human oversight, making responsibility for harmful outcomes difficult to localize among developers, deployers, and operators. Beyond unintentional failures, agentic AI systems are also vulnerable to deliberate manipulation, including data poisoning\cite{alber2025medical} and backdoor attacks \cite{10905032,11045703} that embed hidden behaviors into perception or memory components and later influence autonomous decision-making. Deng \textit{et al.} \cite{10.1145/3716628} survey security risks specific to agentic AI, highlighting how such vulnerabilities can be exploited through adversarial inputs, system-level attacks, and compromised interfaces. In practice, failures in bias mitigation, responsibility attribution, and safety are not exceptional cases but recurring risks inherent to autonomous agent deployment.

\subsubsection{SAD (Societal, Adoption, Design)}
The societal aspects of agentic AI become most evident when autonomous systems are designed and deployed within real social and organizational contexts. Yet these dimensions remain only sparsely examined in existing research. Agentic AI does not merely introduce technical change; it reshapes decision-making practices, redefines participation, and alters how legitimacy and trust are established. Tiwari \textit{et al.} \cite{tiwari2025conceptualising} demonstrate that deployments of agentic urban AI expose value conflicts and ethical pluralism, even in successful applications such as transportation, energy management, and public safety. Their analysis of pilot projects in cities including Singapore and Riyadh shows that technical effectiveness alone is insufficient, and that governance approaches must incorporate participatory mechanisms, value learning, and ethical oversight to sustain public trust. Challenges related to adoption further reflect this socio-technical complexity. Abedin \textit{et al.} \cite{abedin2022designing} argue that integrating agentic AI into organizational settings requires changes to institutional roles, accountability structures, and workplace norms, rather than the simple introduction of new tools. From a design perspective, these challenges translate into the need to balance autonomous system behavior with meaningful human involvement. Te'eni \textit{et al.} \cite{te2025takes} emphasize that effective human–AI systems rely on feedback and control mechanisms that preserve human judgment while allowing agents to learn and adapt. The SAD perspective therefore highlights that societal impact, adoption dynamics, and design choices are tightly interlinked, and that the effectiveness of agentic AI depends on its alignment with social expectations and institutional contexts as much as on its technical capabilities. 

\subsection{Agentic AI and Society}\label{subsec:AAIandSociety}
The agentic AI architectural loop highlights a bidirectional relationship between agentic AI and society, distinguishing between Agentic AI for Society and Agentic AI on Society, and situating agentic systems within broader sociological contexts.
\begin{table*}[]
\centering
\caption{Agentic AI and Society, distinguishing between Agentic AI for Society and Agentic AI on Society, capturing both intended societal use and broader societal impact.}
\label{tab:AISoc}
\rotatebox{90}{\resizebox{!}{.19\paperheight}{%
\begin{tabular}{ccllll}
\hline
\multicolumn{1}{|c|}{\multirow{20}{*}{\rotatebox{270}{\textbf{Agentic AI for Society}}}} &
  \multicolumn{1}{c|}{\textbf{Issues}} &
  \multicolumn{1}{c|}{\textbf{Issue Definition}} &
  \multicolumn{1}{c|}{\textbf{Impact on Society}} &
  \multicolumn{1}{c|}{\textbf{Challenges}} &
  \multicolumn{1}{c|}{\textbf{Potential Solutions}} \\ \cline{2-6} 
\multicolumn{1}{|c|}{} &
  \multicolumn{1}{c|}{\textbf{\begin{tabular}[c]{@{}c@{}}Increase\\ Productivity\end{tabular}}} &
  \multicolumn{1}{l|}{\begin{tabular}[c]{@{}l@{}}Agentic AI can partially or fully offload \\ complex operational tasks from humans.\end{tabular}} &
  \multicolumn{1}{l|}{\begin{tabular}[c]{@{}l@{}}Shifts the value of human expertise and transforms \\ work structures toward automation-centric \\ economies.\end{tabular}} &
  \multicolumn{1}{l|}{\begin{tabular}[c]{@{}l@{}}Job displacement, economic inequality, and \\ uncertainty about future labor organization.\end{tabular}} &
  \multicolumn{1}{l|}{\begin{tabular}[c]{@{}l@{}}Develop adaptive labor policies, reskilling \\ initiatives, and human–AI collaboration \\ models.\end{tabular}} \\ \cline{2-6} 
\multicolumn{1}{|c|}{} &
  \multicolumn{1}{c|}{\textbf{\begin{tabular}[c]{@{}c@{}}Enhance\\ Sustainability\end{tabular}}} &
  \multicolumn{1}{l|}{\begin{tabular}[c]{@{}l@{}}Agentic AI autonomously manages and \\ optimizes complex ecological data.\end{tabular}} &
  \multicolumn{1}{l|}{\begin{tabular}[c]{@{}l@{}}Supports sustainable development and efficient \\ resource utilization.\end{tabular}} &
  \multicolumn{1}{l|}{\begin{tabular}[c]{@{}l@{}}High computational costs and energy \\ consumption contribute to carbon emissions.\end{tabular}} &
  \multicolumn{1}{l|}{\begin{tabular}[c]{@{}l@{}}Invest in green AI research, energy-efficient\\ computing, and carbon offset standards for \\ AI systems.\end{tabular}} \\ \cline{2-6} 
\multicolumn{1}{|c|}{} &
  \multicolumn{1}{c|}{\textbf{\begin{tabular}[c]{@{}c@{}}Personalized\\ Education\end{tabular}}} &
  \multicolumn{1}{l|}{\begin{tabular}[c]{@{}l@{}}Agentic AI provides adaptive and \\ individualized learning experiences.\end{tabular}} &
  \multicolumn{1}{l|}{\begin{tabular}[c]{@{}l@{}}Enhances educational accessibility and \\ personalization.\end{tabular}} &
  \multicolumn{1}{l|}{\begin{tabular}[c]{@{}l@{}}Undermine human critical thinking and \\ academic integrity.\end{tabular}} &
  \multicolumn{1}{l|}{\begin{tabular}[c]{@{}l@{}}Promote AI literacy, embed human oversight\\ in educational AI design, and balance \\ human–machine co-learning.\end{tabular}} \\ \cline{2-6} 
\multicolumn{1}{|c|}{} &
  \multicolumn{1}{c|}{\textbf{\begin{tabular}[c]{@{}c@{}}Knowledge\\ Production\end{tabular}}} &
  \multicolumn{1}{l|}{\begin{tabular}[c]{@{}l@{}}Agentic AI autonomously generates, \\ organizes, and disseminates information.\end{tabular}} &
  \multicolumn{1}{l|}{Expands access to knowledge.} &
  \multicolumn{1}{l|}{\begin{tabular}[c]{@{}l@{}}Ethical dilemmas, misinformation, and \\ intellectual property ambiguity.\end{tabular}} &
  \multicolumn{1}{l|}{\begin{tabular}[c]{@{}l@{}}Establish robust governance, transparency\\ standards, and clear attribution frameworks.\end{tabular}} \\ \cline{2-6} 
\multicolumn{1}{|c|}{} &
  \multicolumn{1}{c|}{\textbf{\begin{tabular}[c]{@{}c@{}}Social\\ Relations\end{tabular}}} &
  \multicolumn{1}{l|}{\begin{tabular}[c]{@{}l@{}}Agentic AI systems learn from human \\ social contexts to simulate empathy \\ and influence behaviors.\end{tabular}} &
  \multicolumn{1}{l|}{Alters interpersonal relationships and social norms.} &
  \multicolumn{1}{l|}{\begin{tabular}[c]{@{}l@{}}Trust deficits, emotional dependency, and \\ security risks.\end{tabular}} &
  \multicolumn{1}{l|}{\begin{tabular}[c]{@{}l@{}}Implement alignment frameworks that allow\\ real-time adjustment of AI behavior to \\ ethical and social cues.\end{tabular}} \\ \cline{2-6} 
\multicolumn{1}{|c|}{} &
  \multicolumn{1}{c|}{\textbf{\begin{tabular}[c]{@{}c@{}}Enhance\\ Automation\end{tabular}}} &
  \multicolumn{1}{l|}{\begin{tabular}[c]{@{}l@{}}Exhibits proactivity, adaptability, and \\ goal-oriented optimization of actions.\end{tabular}} &
  \multicolumn{1}{l|}{\begin{tabular}[c]{@{}l@{}}Transforms organizational efficiency and \\ decision-making speed across sectors.\end{tabular}} &
  \multicolumn{1}{l|}{\begin{tabular}[c]{@{}l@{}}Unclear boundaries of autonomy and \\ accountability in AI actions.\end{tabular}} &
  \multicolumn{1}{l|}{\begin{tabular}[c]{@{}l@{}}Define limits of acceptable autonomy through\\ governance and ethical oversight mechanisms.\end{tabular}} \\ \cline{2-6} 
\multicolumn{1}{|c|}{} &
  \multicolumn{1}{c|}{\textbf{\begin{tabular}[c]{@{}c@{}}Enhance\\ Resilience\end{tabular}}} &
  \multicolumn{1}{l|}{\begin{tabular}[c]{@{}l@{}}Agentic AI continuously refines its \\ strategies based on new data and \\ evolving environmental conditions.\end{tabular}} &
  \multicolumn{1}{l|}{\begin{tabular}[c]{@{}l@{}}Increases effectiveness in unpredictable or rapidly \\ changing scenarios, strengthening systemic \\ adaptability.\end{tabular}} &
  \multicolumn{1}{l|}{\begin{tabular}[c]{@{}l@{}}Potential misalignment with long-term \\ human values and strategic goals.\end{tabular}} &
  \multicolumn{1}{l|}{\begin{tabular}[c]{@{}l@{}}Develop adaptive governance models and \\ multi-layered oversight systems ensuring \\ value alignment under dynamic conditions.\end{tabular}} \\ \hline
\multicolumn{6}{l}{} \\
\multicolumn{1}{l}{} &
  \multicolumn{1}{l}{} &
   &
   &
   &
   \\ \hline
\multicolumn{1}{|c|}{\multirow{15}{*}{\rotatebox{270}{\textbf{Agentic AI on Society}}}} &
  \multicolumn{1}{c|}{\textbf{Societal Issue}} &
  \multicolumn{1}{c|}{\textbf{Definition}} &
  \multicolumn{1}{c|}{\textbf{Impact on Agentic AI}} &
  \multicolumn{1}{c|}{\textbf{Challenges for AI Development}} &
  \multicolumn{1}{c|}{\textbf{Socio-Technical Solutions}} \\ \cline{2-6} 
\multicolumn{1}{|c|}{} &
  \multicolumn{1}{c|}{\textbf{\begin{tabular}[c]{@{}c@{}}Legal and\\ Regulatory \\ Frameworks\end{tabular}}} &
  \multicolumn{1}{l|}{\begin{tabular}[c]{@{}l@{}}Laws determine permissible AI \\ behavior and autonomy.\end{tabular}} &
  \multicolumn{1}{l|}{\begin{tabular}[c]{@{}l@{}}Constrains goal-setting, data usage, and decision \\ authority of Agentic AI systems.\end{tabular}} &
  \multicolumn{1}{l|}{Risk of non-compliance and misuse of AI.} &
  \multicolumn{1}{l|}{\begin{tabular}[c]{@{}l@{}}Develop global AI governance and adaptive \\ regulatory standards.\end{tabular}} \\ \cline{2-6} 
\multicolumn{1}{|c|}{} &
  \multicolumn{1}{c|}{\textbf{\begin{tabular}[c]{@{}c@{}}Ethical and \\ Cultural Norms\end{tabular}}} &
  \multicolumn{1}{l|}{\begin{tabular}[c]{@{}l@{}}Societal values shape what is \\ considered acceptable AI conduct.\end{tabular}} &
  \multicolumn{1}{l|}{\begin{tabular}[c]{@{}l@{}}Influences interaction patterns and decision \\ priorities.\end{tabular}} &
  \multicolumn{1}{l|}{\begin{tabular}[c]{@{}l@{}}Cultural bias and ethical pluralism complicate \\ universal ethical design and deployment.\end{tabular}} &
  \multicolumn{1}{l|}{\begin{tabular}[c]{@{}l@{}}Create context-sensitive ethics frameworks \\ reflecting local norms and global human rights \\ principles.\end{tabular}} \\ \cline{2-6} 
\multicolumn{1}{|c|}{} &
  \multicolumn{1}{c|}{\textbf{\begin{tabular}[c]{@{}c@{}}Data Governance\\ and Privacy \\ Norms\end{tabular}}} &
  \multicolumn{1}{l|}{\begin{tabular}[c]{@{}l@{}}Societal and legal frameworks define \\ how personal and collective data \\ may be accessed, stored, and used.\end{tabular}} &
  \multicolumn{1}{l|}{\begin{tabular}[c]{@{}l@{}}Constrains data availability for AI training and \\ adaptation.\end{tabular}} &
  \multicolumn{1}{l|}{\begin{tabular}[c]{@{}l@{}}Balancing personalization with privacy; \\ limited data diversity reduces model \\ generalizability.\end{tabular}} &
  \multicolumn{1}{l|}{\begin{tabular}[c]{@{}l@{}}Implement transparent, auditable data\\ governance systems emphasizing \\ accountability and privacy preservation.\end{tabular}} \\ \cline{2-6} 
\multicolumn{1}{|c|}{} &
  \multicolumn{1}{c|}{\textbf{Public Trust}} &
  \multicolumn{1}{l|}{\begin{tabular}[c]{@{}l@{}}Collective attitudes toward AI safety, \\ ethical reliability, and legitimacy of \\ autonomous systems.\end{tabular}} &
  \multicolumn{1}{l|}{\begin{tabular}[c]{@{}l@{}}Influences AI adoption, autonomy level, and public\\ acceptance of Agentic decision-making.\end{tabular}} &
  \multicolumn{1}{l|}{\begin{tabular}[c]{@{}l@{}}Low trust leads to underuse; overtrust fosters \\ complacency and misuse.\end{tabular}} &
  \multicolumn{1}{l|}{\begin{tabular}[c]{@{}l@{}}Enhance transparency, participatory \\ governance, and ethical communication \\ strategies.\end{tabular}} \\ \cline{2-6} 
\multicolumn{1}{|c|}{} &
  \multicolumn{1}{c|}{\textbf{\begin{tabular}[c]{@{}c@{}}Control and \\ Accountability \\ Structures\end{tabular}}} &
  \multicolumn{1}{l|}{\begin{tabular}[c]{@{}l@{}}Institutional and corporate contexts \\ define who controls AI systems and \\ for what purposes.\end{tabular}} &
  \multicolumn{1}{l|}{\begin{tabular}[c]{@{}l@{}}Shapes AI’s operational goals and capacity for \\ independent reasoning.\end{tabular}} &
  \multicolumn{1}{l|}{Risk of misuse for manipulation.} &
  \multicolumn{1}{l|}{\begin{tabular}[c]{@{}l@{}}Establish independent oversight, ensure \\ algorithmic auditability, and promote distributed \\ control mechanisms.\end{tabular}} \\ \hline

\end{tabular}
}}
\end{table*}

\subsubsection{Agentic AI for Society}
Agentic AI holds significant potential to generate societal value by enhancing productivity, supporting sustainability goals, extending human capabilities, and broadening access to knowledge across domains. Through the automation of routine tasks and the augmentation of complex decision-making, these systems are already reshaping labor structures, organizational practices, and modes of social participation~\cite{murugesan2025rise,shavit2023practices}, as summarized in Table~\ref{tab:AISoc}. Agentic AI's increasing integration into economic and social systems introduces structural challenges related to employment transitions, skills redistribution, and unequal access to education and training. Without sustained investment in reskilling and inclusive capacity-building, the benefits of agentic AI risk being unevenly distributed, reinforcing existing social and economic disparities rather than alleviating them.

In addition to labor-related implications, agentic AI introduces broader systemic risks associated with autonomous operation in social environments. These include disruptions to institutional processes, unintended consequences arising from misalignment with human intent, and the potential for misuse in contexts such as manipulation, misinformation, and social destabilization. As agentic systems increasingly learn from and act within social contexts, their influence on collective behavior and social norms becomes more pronounced. Addressing these challenges requires governance approaches that explicitly account for the societal embedding of agentic AI and ensure alignment with shared human values and long-term social objectives.

\subsubsection{Agentic AI on Society}
On the other hand, Table.~\ref{tab:AISoc} also illustrates the key societal dimensions that influence the design, operation, and legitimacy of agentic AI systems, along with their corresponding challenges and socio-technical remedies. Legal frameworks delineate permissible AI behaviors and decision authority, while ethical and cultural norms guide acceptable conduct; however, legal and ethical pluralism complicate the establishment of universal regulatory and moral standards~\cite{banchio2024legal}. The issues are similar to those discussed in the MAD and SAD dimensions. Data governance defines how personal and collective data are collected, stored, and utilized, whereas public trust determines the extent to which AI is accepted and integrated into social and institutional contexts~\cite{murugesan2025rise}. Similarly, control and accountability (in connection with BAD dimension) structures specify who governs AI systems and for what purposes, shaping the distribution of decision-making power and the potential for responsible oversight~\cite{joyce2024sociology}. 
However, persistent issues such as algorithmic bias and data inequities pose significant risks to fairness, inclusion, and transparency as discussed in the BAD dimension. Biases embedded in data or model design can perpetuate systemic discrimination, undermining both public trust and the ethical integrity of Agentic AI. 

\subsubsection{Agentic AI Sociology}
As agentic systems gain the capacity for self-governance, agency, and autonomy, their decisions increasingly intersect with human values, institutional norms, and moral expectations~\cite{biswas2025building}. However, with these advanced capabilities, agentic AI systems are able to act independently and influence social environments by granting non-human entities the capacity to generate, interpret, and even co-produce social data. Such agentic socio-technical systems have the potential to reconstruct or reproduce new forms of social life~\cite{joyce2024sociology}. The integration of autonomous systems into human society challenges conventional notions of accountability, governance, and social participation, necessitating a deeper understanding of how agentic AI reshapes social structures and moral frameworks. Consequently, critical questions arise: \textit{What are the boundaries and implications of social inclusion when non-human agents become participants in social processes?  Who possesses the authority to define norms? What risks accompany their independent decision-making? Should AI be allowed to make moral or emotional judgments? Could machines one day develop their own social norms?} 
This trajectory suggests an emerging form of distributed agency, where humans, artificial agents, and institutions jointly participate in decision-making processes, complicating traditional understandings of social order and responsibility.

\subsection{Ethical aspects}
The ethical implications of agentic AI are central to its development and deployment. As these systems acquire greater autonomy, they extend beyond technical challenges to actors whose decisions can affect individuals, organizations, and society at large. Core ethical concerns include fairness, accountability, transparency, privacy, and safety, all of which must be addressed to ensure that agentic AI systems remain aligned with societal values and ethical norms. As agentic AI systems continue to evolve, sustained attention to ethical integration will remain essential for aligning technical capability with societal expectations.

A major ethical challenge lies in bias in autonomous decision-making \cite{mezgar2022ethics}. Because agentic AI systems learn from large-scale datasets, biases embedded in historical or institutional data can be reproduced and amplified through autonomous reasoning and action. In application domains such as hiring, law enforcement, and healthcare, such biases may disproportionately affect marginalized groups and reinforce existing inequalities. Addressing this requires careful monitoring and intervention to ensure that AI systems are not only transparent but also capable of identifying and mitigating bias in their decision processes. Accountability is another ethical challenge \cite{mittelstadt2019principles}. When decisions are generated and executed autonomously, questions arise regarding the respective roles of developers, deployers, users, and the system itself. Addressing this ambiguity requires clear accountability frameworks that support traceability, auditability, and post hoc analysis of autonomous decisions, particularly in high-stakes or safety-critical settings.

In addition, the transparency of decision-making processes in agentic AI is a critical ethical consideration \cite{wachter2017transparent}. Many AI systems, especially those based on deep learning models, operate as "black boxes," meaning that their decision-making processes are not easily understood by humans~\cite{kaplan2024unified}. This lack of transparency can lead to mistrust and concerns over how decisions are made, particularly when these decisions impact people's lives. To address these concerns, it is important to prioritize explainability\cite{papagni2023artificial}, enabling users and stakeholders to understand how and why a particular decision was made by an AI system. Privacy is another crucial ethical issue \cite{tzanis2025agentic}. Agentic AI systems rely heavily on data to function, often requiring access to large amounts of personal information. The collection, use, and storage of this data raise concerns over privacy, particularly in terms of informed consent and the protection of sensitive information. Ethical AI systems must respect individuals' privacy rights, ensuring that personal data is handled responsibly and in compliance with relevant regulations. Additionally, the design of such systems should include mechanisms that allow individuals to control their data and opt-out of being monitored or tracked. Finally, the safety and security of autonomous agentic AI systems are paramount \cite{gonzalez2025ai}. Given their increasing autonomy, these systems must be designed with robust safeguards to prevent malfunction or misuse. For example, in critical areas such as healthcare or autonomous transportation, even a small error or failure could have catastrophic consequences. Maintaining system reliability requires fail-safe mechanisms, real-time monitoring, and continuous updates to mitigate vulnerabilities over time.

\subsection{Implications and Challenges}\label{subsec:ImplChal}
The integration of agentic AI across application domains introduces a range of ethical and societal challenges that must be addressed to support responsible deployment. As agentic AI systems become more autonomous and complex, understanding and mitigating associated risks is essential to ensure their safe and trustworthy use. To clarify how ethical and societal concerns relate to technical design, Table~\ref{tab:MAD_BAD_SAD} maps the core technical modules of agentic AI onto the MAD–BAD–SAD perspective. These challenges highlight that the successful integration of agentic AI depends not only on technical capability, but also on sustained attention to societal impact, value alignment, governance, and safety. 

\begin{table}[t]
\centering
\caption{Mapping of Agentic AI Technical Modules to their ethical and societal challenges under the MAD–BAD–SAD framework.}\label{tab:MAD_BAD_SAD}
\renewcommand{\arraystretch}{1.3}
\setlength{\tabcolsep}{3pt}
\resizebox{!}{.24\textwidth}{%
\begin{tabular}{>{\raggedright\arraybackslash}p{1.8cm}
                >{\raggedright\arraybackslash}p{4cm}
                >{\raggedright\arraybackslash}p{4cm}
                >{\raggedright\arraybackslash}p{4cm}}
\toprule
\rowcolor[HTML]{FFFFFF}
\textbf{Technical} & \textbf{MAD} & \textbf{BAD} & \textbf{SAD} \\
\midrule

\rowcolor[HTML]{F7F7F7}
\textbf{Perception} &
Dilemma: Surveillance vs. utility. &
Bias: Encoding societal biases from data; Danger: Adversarial attacks on sensors. &
Design: Privacy-by-design; User consent mechanisms. \\

\textbf{Cognition} &
Autonomy: Self-directed reasoning; Moral reasoning capability. &
Accountability: “Black-box” problem obscures responsibility. &
Adoption: Trust in incomprehensible decisions. \\

\rowcolor[HTML]{F7F7F7}
\textbf{Planning} &
Dilemma: Optimization for goals at the expense of ethics. &
Danger: Generating harmful or unexpected plans. &
Societal: Impact on jobs and economic structures. \\

\textbf{Execution} &
Morality: Carrying out a morally questionable action. &
Accountability: Direct cause of harm; Liability. &
Design: Need for human-in-the-loop override mechanisms. \\

\bottomrule
\end{tabular}
}
\end{table}

One of the primary implications of agentic AI is its potential to disrupt existing systems and structures\cite{10.1145/3716628}. As these systems gain greater autonomy, they are likely to reshape industries by automating tasks that were once handled by humans. In sectors like transportation, healthcare, and manufacturing, this disruption could lead to significant efficiency gains, cost reductions, and new forms of innovation. However, this transformation also carries the risk of job displacement, particularly for those in roles that are most susceptible to automation. Addressing this challenge will require a coordinated effort to develop workforce retraining programs and policies that promote economic adaptability in the face of technological change. A major challenge facing the adoption of Agentic AI is ensuring its alignment with human values and societal goals \cite{pathak2024ai}. They must be capable of making decisions that reflect ethical principles and societal norms. Ensuring that agentic AI systems are not only technically effective but also ethically sound requires ongoing research into value alignment, trustworthiness, and fairness. This is particularly important in high-stakes areas, such as healthcare and criminal justice \cite{ajmani2024data}, where AI decisions can have a profound impact on people's lives. Developers must grapple with the challenge of designing AI systems that can navigate complex moral dilemmas while being transparent and accountable for their decisions.

Governance and regulation represent a critical challenge for the development and deployment of agentic AI systems \cite{buhmann2021towards}. Existing legal and regulatory frameworks are often ill-equipped to address the risks introduced by highly autonomous systems whose actions may evolve beyond direct human control. This gap raises unresolved questions concerning accountability, transparency, and safety, particularly when agentic AI operates across sectors and jurisdictions. Effective governance, therefore, requires not only updated legal instruments but also coordination across national boundaries, as the impacts of agentic AI systems routinely extend beyond localized regulatory regimes. Agentic AI agents must remain resilient to adversarial threats, including cyberattacks, as well as to unanticipated system failures that may arise during long-term operation. Addressing these risks demands the integration of security mechanisms, continuous monitoring, and fail-safe controls capable of constraining harmful behavior.  Also, there is a need for an anticipatory governance framework that embeds human values throughout the system lifecycle and supports sustainable policymaking \cite{garg2025designing,liu2025agentic}. The opacity of many learning-based models limits users’ ability to understand and contest autonomous decisions, undermining trust and impeding adoption.

\section{Applications and Case studies} \label{applications}
This section discusses various societal applications and case studies of agentic AI, illustrating how autonomous systems are designed, deployed, and governed in real-world settings.

\subsection{Healthcare and Social Services}
In healthcare, agentic AI systems are being explored to support care coordination, clinical workflow management, diagnosis, and adaptive treatment planning \cite{karunanayake2025healthcare}. These systems can reason over multimodel medical data, clinical guidelines, and evolving patient states, enabling continuous support for chronic disease management and personalized care delivery \cite{jiang2025large,acharya2025agentic}. Unlike traditional decision-support tools, agentic AI can initiate actions, adjust plans in response to new information, and coordinate across multiple stakeholders within clinical institutions. In social services \cite{ALOTAIBI2024273}, agentic AI has shown potential in domains such as elderly care \cite{chen2023agent}, mental health support, and community assistance. Autonomous agents can engage in long-term interaction, monitor behavioral and environmental signals, and coordinate interventions across caregivers, healthcare providers, and social organizations \cite{khalil2025redefining}. These capabilities are particularly relevant in aging societies and resource-constrained settings, where sustained human support is difficult to provide at scale. 
The deployment of agentic AI in these domains also introduces substantial socio-technical challenges in these fields. Clinical and social decisions are often value-laden, context-dependent, and high-stakes, requiring careful calibration of autonomy and human oversight. Biases in data or decision policies may disproportionately affect vulnerable populations, while opaque reasoning processes can undermine professional accountability and patient trust \cite{mittelstadt2019principles}. Privacy and data governance concerns are further intensified by the sensitivity of health and social data, necessitating robust safeguards and regulatory compliance. 

\subsection{Education and Social Behavior Modeling}
LLM–based agentic workflows represent a promising direction for future design and innovation in AI for education, as they can be flexibly applied across educational contexts with varying degrees of autonomy and human–AI interaction through multi-agent collaboration, thereby supporting the continuous evolution of educational technology \cite{dai2024agent4edu}. By enabling natural language processing, LLMs can efficiently address educational tasks and, when embedded in multi-agent systems, generate targeted and specific feedback in the form of suggestions that effectively support the development of learners’ skills while reducing teachers’ instructional workload \cite{dai2024agent4edu}, as exemplified by pedagogical agents shown to positively affect learning outcomes \cite{fountoukidou2019effects}. 
Kamalov \textit{et al.} \cite{kamalov2025evolution} analyze AI agentic workflows in education around four core paradigms: reflection, planning, tool use, and multi-agent collaboration. Collectively, these paradigms enable agents to analyze past performance to identify errors or areas for improvement and refine their future behavior, decompose complex goals, leverage external tools (e.g., calculators, code execution, web search, APIs) to augment their capabilities, and coordinate specialized roles to provide more adaptive and consistent educational support. This enables AI agents to function as reasoning partners while also improving sustainability by reducing computational redundancy through collaborative agent frameworks and targeted task decomposition \cite{kamalov2025evolution}. However, there is a need to prioritize explainable AI to ensure transparency for educators and learners, while also developing adaptive AI agents that can address diverse learner needs, languages, cultural backgrounds, and learning preferences \cite{kamalov2025evolution}.

\subsection{Smart and Sustainable Cities and Urban Transitions}
Agentic AI is increasingly recognized as a critical enabler of smart and sustainable urban transitions, where cities seek to reconfigure mobility, energy, governance, and public services in response to environmental pressures and growing societal complexity \city{tiwari2025conceptualising}. Unlike conventional smart city technologies that rely on centralized analytics or static optimization, agentic AI systems operate as autonomous decision-making entities capable of sensing urban dynamics, reasoning across heterogeneous data streams, and coordinating actions across distributed infrastructures and institutional actors. These characteristics make agentic AI particularly suited to managing long-term urban transitions that involve uncertainty, competing objectives, and continuous adaptation.

Existing studies demonstrate that agentic and multi-agent AI systems provide effective mechanisms for adaptive urban infrastructure management. In intelligent transportation systems, autonomous agents have been employed to coordinate traffic signals, mitigate congestion, and manage multimodal mobility flows in real time, yielding improvements in efficiency, robustness, and network-level performance compared to static or centrally controlled approaches \cite{chu2019multi}. Similarly, in urban energy systems, agentic AI has been applied to decentralized demand response, renewable energy integration, and building-level control, enabling coordination across distributed assets and contributing to low-carbon and resilient energy transitions \cite{5675295,mittal2019agent}. These studies illustrate how agent-based control architectures support coordination across spatial and temporal scales that are difficult to manage using conventional optimization or rule-based frameworks.
Autonomous agents have been used to assist emergency response coordination, dynamic resource allocation, and scenario-based policy analysis in complex urban environments \cite{chen2023agent}. Within the broader context of sustainability transitions, such systems enable cross-sector coordination across transportation, energy, housing, and public services, supporting adaptive planning processes that account for long-term uncertainty and competing objectives \cite{tiwari2025conceptualising}.

\subsection{Resource-Constrained Networking and Communication}
In networking and communication systems, such as wireless sensor networks, Internet of Things, or ad-hoc networks-resources like energy, bandwidth, processing power, and memory are often limited. Also, these systems involve dynamic topologies (nodes joining/leaving, changing links), latency, and reliability requirements. Not only the resource-constraints are dynamic \cite{vaishnav2023multi, vaishnav2025adaptive}, but the priority between different resources and objectives can also dynamically vary with time \cite{vaishnav2025dynamic}. Agentic AI can make significant contributions to the highly constrained and dynamic nature of such environments. 

\subsubsection{Agentic AI in networking}
Across the existing literature, \emph{agentic AI} mainly refers to many semi-autonomous decision-makers operating within the network. These decision-makers can be Unmanned Aerial Vehicles (UAVs)\cite{10934003,10144626}, base stations, access points, road-side units (RSUs)\cite{10578028}, or end-devices. Each one runs a learned policy and reacts to its local conditions, such as queue backlog, channel quality, and remaining energy. Together, they manage scarce resources like spectrum, CPU cycles, and battery. Instead of using fixed scheduling or offloading rules, the agents are trained using multi-agent reinforcement learning (MARL)\cite{11098525}. Training is often done centrally, although some works use federated learning to keep data local. The goal is usually to balance several competing objectives: low latency, low energy consumption, fresh information (AoI), fairness, and sometimes tail reliability (e.g., 95th-percentile delay). Coordination between agents is typically kept simple. In many cases, it is implicit, happening through shared wireless interference and shared critics during training. In other cases, it is made lightweight through federated aggregation or compressed message exchange. This is consistent with the practical constraints of resource-constrained networks.

\begin{table*}[t]
\centering
\caption{Compact overview of agentic AI formulations for resource-constrained networking and communication.}
\label{tab:agentic-ai-resource-constrained-networking}
\resizebox{!}{.325\textwidth}{%
\begin{tabular}{|r|p{4.5cm}|p{4.3cm}|p{4.5cm}|p{4.0cm}|}
\toprule
\rowcolor[HTML]{F7F7F7}
\textbf{Ref} & \textbf{Agentic AI idea (who are the agents?)} & \textbf{Main constraint focus} & \textbf{Coordination / learning style} & \textbf{Application} \\
\midrule
\textbf{\cite{10091485}} &
Phones + edge server as multi-agent RL for offloading \& CPU/bandwidth &
QoS (task success, delay, fairness) under limited edge CPU and radio &
Centralized training, local execution; SAC with heuristic guidance &
Single-cell MEC for many mobiles \\
\hline
\textbf{\cite{10737406}} &
Edge caches as MARL agents deciding when to refresh content &
Data freshness (AoI) vs.\ update cost; penalties for AoI violations &
MADDPG with global critic option (MACU-GC) to cut training cost &
IIoT edge caching for mobile robots \\
\hline
\textbf{\cite{10934003}} &
Many UAVs + UEs in a multi-agent actor--critic game &
Hard limits on UAV motion, power, CPU; target energy efficiency &
Modular actor--critic (DSPAC-MN); centralized training, overhead analyzed &
Multi-UAV MEC for task offloading \\
\hline
\textbf{\cite{10578028}} &
RSUs/vehicles as agents for offloading and resource allocation &
Task deadlines, delay and energy budgets in vehicular MEC &
Hybrid centralized/distributed MARL; federated learning for privacy &
IoV MEC with many RSUs and cars \\
\hline
\textbf{\cite{11098525}} &
Each Wi-Fi station as an independent RL agent for RU + power &
Per-user AoI targets and power limits in UL-OFDMA &
Independent PPO agents; fully decentralized, implicit coordination via interference &
AoI-oriented Wi-Fi for IoT traffic \\
\hline
\textbf{\cite{10179935}} &
UAV-MEC entities as agents for when/where to process tasks &
Average AoI and ``recent tasks'' under capacity constraints &
Federated multi-agent RL; agents share parameters, keep data local &
UAV-assisted MEC for fresh sensing data \\
\hline
\textbf{\cite{10533447}} &
UAVs in HAP+UAV NTN as RL agents for route \& resource control &
AoI under uncertain CSI, power, CPU, trajectory limits &
MADDPG plus federated RL variant; CTDE with robustness to CSI errors &
Aerial computing in non-terrestrial networks \\
\hline
\textbf{\cite{10144626}} &
UAVs/edge servers as federated RL agents for medical offloading &
Latency \& energy subject to IoMT QoS and coverage gaps &
Multi-Agent Federated RL (MAFRL); local training, global aggregation &
UAV-enabled Internet of Medical Things \\
\hline
\textbf{\cite{10485479}} &
MEC devices as agents that also learn when/what to communicate &
Limited energy and inter-device bandwidth; QoE (delay+energy) &
Scheduled multi-agent DRL with Top-K message passing; CTDE &
Resource-constrained MEC with weak connectivity \\
\bottomrule
\end{tabular}%
}
\end{table*}

\subsubsection{Agentic AI formulations for Networking}
A consistent design choice across the studies is how the environment is decomposed into decision-making units. In several Mobile edge computing (MEC)\cite{ 10485479} and aerial networking settings, the \emph{infrastructure itself} becomes the agent: UAVs and aerial platforms act as mobile edge servers that jointly decide trajectory, computation allocation, and offloading policies \cite{10934003, 10179935, 10533447, 10144626}. In other cases, coordination is pushed closer to the radio layer, where base stations, RSUs, or spectrum schedulers behave as high-level controllers operating under tight reliability or deadline constraints \cite{10091485, 10578028, 10485479}. At the opposite end, some works explicitly model \emph{end-devices} as independent agents. For example, each Wi-Fi station in UL-OFDMA chooses resource units (RUs) and transmit power to meet AoI targets under local power limits \cite{11098525}. In the most communication-constrained cases, agents are not only making offloading decisions but also learning \emph{when and what to communicate}, treating messaging itself as a constrained action \cite{10485479}. Overall, the ``agentic'' label here is quite literal: distributed entities with local state, their own policy, and interactions mediated by the shared wireless environment.

\subsubsection{Network constrains in Agentic AI}
The constraints in networking is a mix of hard physical limits and softer service guarantees. Hard limits—such as transmit power, CPU cycles, bandwidth, or UAV flight dynamics—are usually enforced by bounding the action space, adding feasibility checks, or embedding resource budgets directly into the simulator environment \cite{10091485, 10737406, 10934003, 10179935, 10533447, 10144626, 10485479}. In contrast, service-oriented constraints (deadlines, AoI targets, fairness requirements, Ultra Reliable and Low Latency Communications (URLLC)-style reliability) are more commonly handled through reward engineering: violations are penalized, and the agent learns policies that trade off constraint adherence against performance. This shows up clearly in healthcare IoT, where blocking probability and tail delay (e.g., 95th-percentile delay) are key reliability indicators, and in AoI-sensitive scenarios, where freshness violations become explicit penalty terms (e.g., caching, Wi-Fi uplink, and aerial sensing) \cite{10737406, 11098525, 10179935, 10533447}. Importantly, the emphasis is rarely on formal constrained-RL guarantees; instead, the constraint-handling is engineering-oriented, aiming for implementable policies under realistic system limits.

\subsubsection{Coordination and learning: central, federated, or fully decentralized}
Coordination mechanisms track the communication realities of the target networks. For continuous-action control problems (e.g., power and CPU allocation, UAV mobility), centralized training with decentralized execution (CTDE) is a common compromise: critics have access to global state during training, while agents act locally at run-time \cite{10091485, 10737406, 10934003, 10179935, 10533447, 10144626, 10485479}. Where privacy, scalability, or data locality matters, federated multi-agent RL becomes attractive, particularly in healthcare and IoMT, as well as UAV-MEC and Non-Terrestrial Networks (NTN) contexts \cite{10578028, 10179935, 10533447, 10144626}. At the same time, some scenarios deliberately avoid explicit inter-agent messaging because the system cannot afford it; instead, coordination is implicit through interference patterns and the shared environment dynamics, as in URLLC subnetworks trained with asynchronous updates and Wi-Fi stations trained as independent PPO agents \cite{11098525}. A notable extension is communication-efficient MARL, where agents learn an explicit messaging policy under tight bandwidth constraints, using mechanisms such as gradient sparsification \cite{10742099} and scheduled Top-K message passing \cite{ 10485479}. This direction is particularly relevant for resource-constrained networks because it treats coordination overhead as a first-class system cost rather than an assumption.

\subsubsection{Application flavors in resource-constrained networking}
Although the settings vary, the applications cluster into three recurring categories. First, MEC offloading and resource allocation problems dominate: phones, vehicles, robots, sensors, and UAV servers jointly decide where to compute tasks and how to allocate limited compute/radio resources \cite{10091485, 10934003, 10578028, 10179935, 10533447, 10144626, 10485479}. Second, radio and spectrum management appear in more reliability-driven forms, including URLLC-style scheduling and AoI-aware RU/power control \cite{11098525}. Third, caching and information freshness show up as an AoI-driven control problem, where the agent balances freshness against update cost and backhaul usage \cite{10737406}. Taken together, these case studies suggest a relatively stable template for agentic AI in constrained networks: represent network entities as decision-makers, encode constraints through action feasibility and reward penalties, and choose coordination mechanisms that respect (rather than ignore) the cost of communication.

\subsection{Earth Observation \& Satellite Communications}
\begin{figure}[t]
    \centering
    \includegraphics[width=0.75\textwidth]{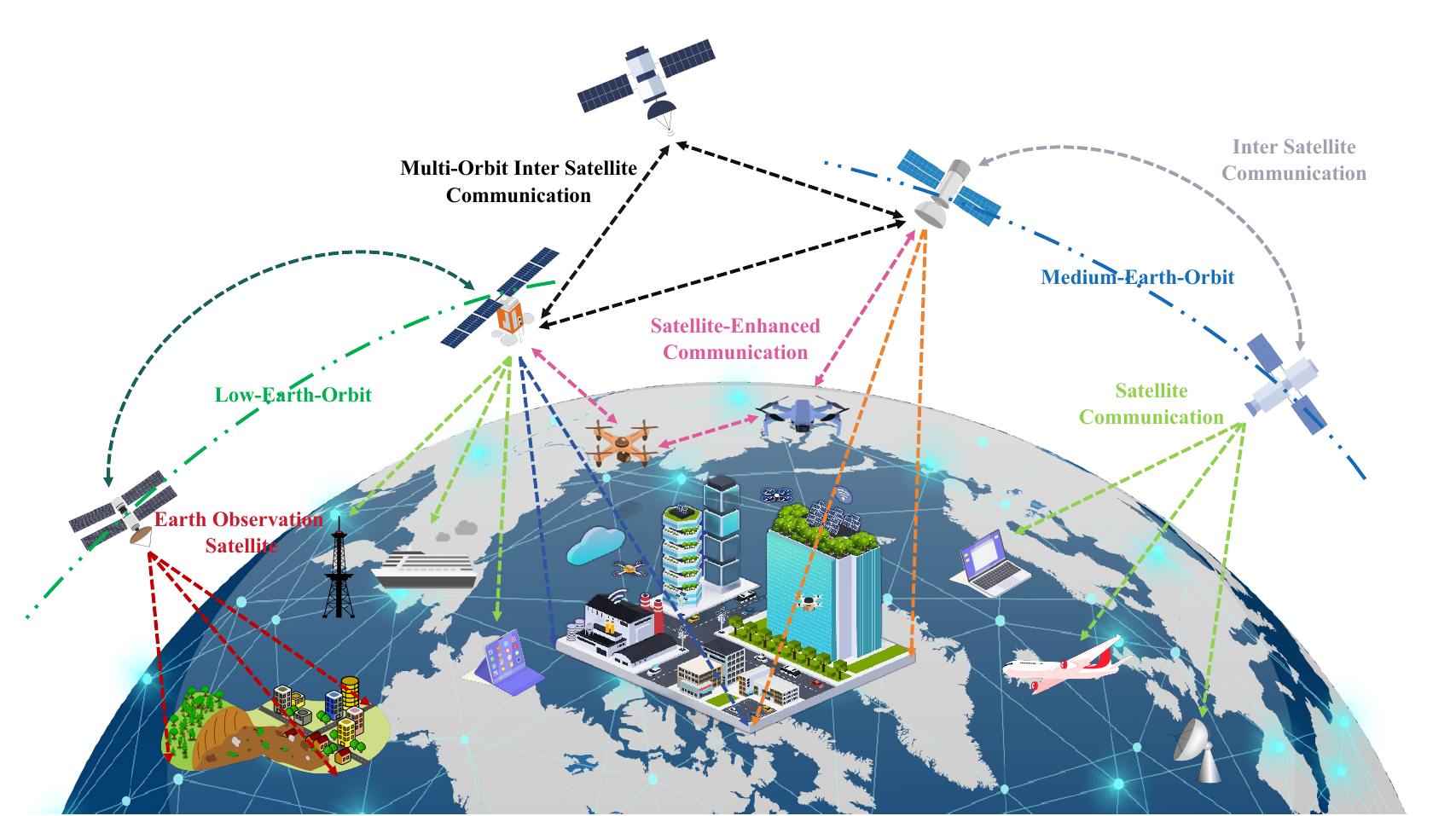}
    \caption{The overall architecture of the satellite communication system.}
    \label{fig:EOS}
\end{figure}

Recent progress in agentic AI systems is fundamentally reshaping Earth Observation (EO) and satellite communications fields, as shown in Fig.~\ref{fig:EOS}.  Agentic AI systems enable proactive sensing strategies, adaptive communication planning, and coordinate multi-satellite operations with minimal human involvement. 


\begin{table*}[t]
\centering
\caption{Comparison of Agentic AI Methods for Earth Observation}\label{tab:EO}
\resizebox{!}{.3\textwidth}{%
\begin{tabular}{r p{2cm} p{5.2cm} p{3.7cm} p{3.7cm}}
\toprule

\textbf{Ref} & \textbf{Method} & \textbf{Technical Approach} & \textbf{Advantages} & \textbf{Limitations} \\
\midrule
\cite{denby2019orbital} & OEC & Leverages onboard computing for data pre-processing and compression to reduce downlink load. & Reduces bandwidth demand; lowers ground processing. & Limited onboard resources; restricted support for complex tasks. \\
\hline
\cite{tao2024known} & Serval & Prioritizes imagery using static/dynamic filtering with distributed satellite-ground processing. & Flexible prioritization; improves responsiveness. & Fixed compression; limited adaptability across scenarios. \\
\hline
\cite{kodan} & Kodan & Filters low-value data and adapts transmission precision for efficient downlink. & Reduces redundant data; increases efficiency. & Depends on data valuation; potential information loss. \\
\hline
\cite{du2025earth+} & Earth+ & Transmits only differences from reference images to minimize bandwidth. & Reduces transmission volume; improves efficiency. & Relies on reference image quality; best for low-change scenes. \\
\hline
\cite{zhang2025satcooper} & SatCooper & DRL-based online algorithm for global, fine-grained LEO–MEO coverage. & Adapts to dynamic environments; improves coverage robustness. & High training cost; requires accurate system modeling. \\
\hline
\cite{li2021service} & Service Placement & Online placement of services on satellite edge nodes for robustness-aware coverage. & Enhances service robustness and coverage. & High complexity; sensitive to resource allocation. \\
\hline
\cite{teng2025efficient} & Knowledge Service Framework & Lightweight onboard models combined with ground knowledge graphs for real-time processing and causal inference. & Efficient and stable; supports complex reasoning; reduces bandwidth usage. & Limited onboard inference; relies on ground knowledge base. \\
\bottomrule
\end{tabular}}
\end{table*}


Within EO, agentic AI capabilities support on-demand tasking of satellite constellations, flexible sensor scheduling, and intelligent selection of observation priorities based on mission objectives or environmental dynamics. As a result, both coverage and data efficiency improve, enabling faster responses to rapidly evolving events such as wildfires, floods, or deforestation.
As summarized in the Table. \ref{tab:EO},
Orbital Edge Computing (OEC) \cite{denby2019orbital} exemplifies this shift by moving computation onboard to ease the downlink bottleneck between satellites and ground stations.
Serval \cite{tao2024known} introduces a filtering-based transmission strategy for latency-critical applications, allowing workloads to be distributed between space and ground.
Kodan \cite{kodan} enhances downlink efficiency by filtering low-value data and adapting data precision, while Earth+ \cite{du2025earth+} minimizes bandwidth consumption by sending only image differences rather than full scenes.
These approaches maximize onboard resource utilization and streamline data delivery.
To further enhance autonomy, SatCooper \cite{zhang2025satcooper} develops a DRL–based online algorithm for fine-grained global coverage under uncertain network conditions across low and medium Earth orbits.
Li \textit{et al.} \cite{li2021service} design an online service placement algorithm that deploys functions on satellite edge computing nodes to ensure robust coverage under resource constraints.
Teng \textit{et al.} \cite{teng2025efficient} present a lightweight knowledge service framework that integrates onboard models for real-time processing with ground-based knowledge graphs for high-level inference and causal reasoning.
Satellite communication networks support a broad range of sensing and communication tasks, as illustrated in Fig. \ref{fig:EOS}.
Yet, inherent limitations, such as low resolution, data gaps, and noise, often degrade the quality and usability of the collected information.
Conventional augmentation methods struggle in low-quality or high-noise scenarios, whereas agentic AI can autonomously enhance data acquisition, analysis, and multimodal fusion even in adverse conditions. Generative AI approaches have demonstrated significant gains in large-scale climate prediction \cite{wang2021precipgan}, GNSS-based positioning \cite{wu2024t}, and satellite data processing \cite{ramirez2025super}. Such intelligence enables resilient, low-latency communication in dynamic contexts including disaster response and global connectivity in remote regions.

\subsection{Industry 5.0 \& Robotics}

Industry 5.0 represents an evolutionary step beyond the automation- and efficiency-driven paradigm of Industry 4.0 by explicitly emphasizing human-centricity, sustainability, and system resilience \cite{nahavandi2019industry,yanytska2025rise}. Rather than pursuing full autonomy and replacing humans, Industry 5.0 promotes the integration of human expertise with intelligent robots and AI agents on the factory floor, which clearly shows socio-technical aspects. For example, Xu \textit{et al.} \cite{xu2021industry} articulate this conceptual shift by contrasting the technology-centric vision of Industry 4.0 with the value-driven and socially aware objectives of Industry 5.0. In this context, recent research has increasingly focused on agent-based and human-centered AI approaches as key enablers of next-generation industrial systems. 
Piccialli \textit{et al.} \cite{piccialli2025agentai} survey autonomous and distributed AI agents in industrial environments, highlighting how Agent AI agents developed for Industry 4.0 extend toward Industry 5.0 requirements. Complementing this perspective, Passalacqua \textit{et al.} \cite{passalacqua2025human} provide a systematic review of human-centered AI in Industry 5.0, demonstrating 67 recent studies on the integration of AI into industrial systems with a focus on human factors. Ethical and value-oriented technology design is further underscored by Longo \textit{et al.} \cite{longo2020value}, who emphasize that societal values must guide the engineering of intelligent industrial systems. From a technological standpoint, early surveys by Maddikunta \textit{et al.} \cite{maddikunta2022industry} identify edge computing, digital twins, collaborative robots, Internet of Everything, blockchain, and 6G as supporting technologies for Industry 5.0. 

Bousdekis \textit{et al.} \cite{bousdekis2019decision} show how cognitive agents can augment decision-making processes in Industry 4.0 manufacturing environments, while Pulikottil \textit{et al.} \cite{pulikottil2023agent} present a comprehensive review of multi-agent systems (MAS) in smart manufacturing, combined with an expert survey to assess industrial adoption.  Practical implementations, such as the agentic AI in a smart factory scheduling system using a multi-agent framework enhanced with digital twins proposed by Siatras \textit{et al.} \cite{siatras2024production}, illustrate the feasibility of these concepts. Furthermore, learning-based approaches are increasingly applied to human–robot collaboration and factory optimization, including RL with human feedback for task allocation \cite{wang2025reinforcement} and MARL for decision-making in smart factories \cite{bahrpeyma2022review}. 
These advances are complemented by studies in collaborative robotics \cite{javaid2021substantial}, flexible manufacturing via agent-based architectures \cite{gross2024agent,nilsson2023customization}, autonomous industrial robotics \cite{fan2025embodied}, and digital twins for agentic decision-making \cite{kusiak2025agentic}, as well as autonomous logistics in factory environments \cite{elfathi2025intelligent}.

\section{Open Challenges and Future Directions} \label{future}

\subsection{Open Challenges}
In addition to the challenges discussed in subsection~\ref{subsec:ImplChal}, several important challenges remain open and need further attention.
\subsubsection{Technical and Architectural Challenges}
Agentic AI systems deployment often depends on heterogeneous hardware stacks that combine edge devices and cloud-based resources. Variations in hardware reliability, energy availability, and communication latency can significantly affect system performance and stability. 
\begin{description}
\item[System Robustness and Safety]  are critically important in high-stakes applications of agentic AI, such as communication systems, drug discovery~\cite{jumper2021highly,stokes2020deep}, and autonomous decision-making. These systems often operate in complex and dynamic environments, where errors are more likely to occur ~\cite{raheem2025agentic}. Such failures can lead to catastrophic consequences, especially when the tolerance for faults is low. Therefore, ensuring robustness and safety in these complex, safety-critical domains \cite{{donta2025resilientdesignactive}} is of paramount importance. Moreover, agentic AI systems are highly data-driven, which makes them particularly vulnerable to adversarial attacks both at the data level (e.g., input perturbations) and the parameter level (e.g., model poisoning~\cite{murugesan2025rise} . Developing methods to maintain stability and reliable performance under such adversarial conditions remains an urgent and open challenge.
\item[Hallucinations and Misinterpretations] Recent progress in LLMs has expanded their representational capacity and broadened their use in autonomous systems, enabling agentic AI to perform reasoning, decision-making, and multi-step task execution. However, hallucination remains a persistent technical limitation, particularly for large, opaque models such as ChatGPT and GPT-4~\cite{sun2025towards,maynez-etal-2020-faithfulness}. In the context of LLMs, hallucination is commonly defined as the generation of content that is linguistically coherent yet factually inaccurate, logically inconsistent, or entirely fabricated. This phenomenon undermines system reliability and poses significant risks in domains where errors have serious consequences, including healthcare, legal decision-making, and scientific analysis. Although the precise causes of hallucination are not fully characterized, prior works ~\cite{piccialli2025agentai,li2024enhancing,lee2025agentic} attributes it to factors such as probabilistic token prediction, gaps or biases in training data, weak grounding in external knowledge, and limited support for explicit reasoning. Reducing hallucination, therefore, remains a central challenge for improving the robustness, interpretability, and epistemic reliability of LLM-based agentic systems.
One promising direction involves integrating external tools and feedback mechanisms into agentic workflows, particularly in emerging environments such as 6G-enabled systems~\cite{ale2025enhancing}. Tool-augmented agents can query external data sources, invoke domain-specific services, and validate intermediate results during execution. This interaction provides a mechanism for grounding model outputs in observable and verifiable information, reducing reliance on unconstrained internal inference. Empirical studies suggest that such grounding can improve contextual accuracy and lower hallucination rates across applications, including healthcare, autonomous control, and smart infrastructure. While tool integration does not eliminate approximation or uncertainty in model outputs, it constrains generation through interaction with external evidence, offering a practical pathway toward more dependable agentic AI behavior.

\end{description}

\subsubsection{Socio-Economic and Organizational Challenges}
The deployment of agentic AI systems introduces a range of socio-economic and organizational challenges that shape how such technologies are adopted, governed, and sustained in practice. One of the most visible concerns is job displacement and workforce transformation, as agentic AI automates not only routine tasks but also decision-making and coordination functions traditionally performed by skilled professionals. While these systems can augment productivity and enable new forms of work, they also risk displacing roles, restructuring occupational hierarchies, and widening skill gaps if reskilling and transition pathways are not adequately supported. Also, organizations often encounter cultural and institutional resistance when introducing agentic AI. Autonomous systems challenge established norms of expertise, authority, and accountability, leading to skepticism or pushback from employees who may perceive a loss of control or professional identity. Such resistance is frequently compounded by misalignment between technical capabilities and organizational values.

Unrealistic expectations and return-on-investment pressures further complicate adoption. Agentic AI is often promoted as a solution capable of rapid efficiency gains and cost reduction, creating short-term performance expectations that are difficult to meet in complex, real-world settings. When benefits fail to materialize at the anticipated pace, organizations may prematurely abandon systems or deploy them inappropriately, increasing operational risk. These pressures are exacerbated by the substantial costs associated with developing, maintaining, and governing agentic AI, including infrastructure investment, data management, and ongoing oversight. Finally, workflow integration challenges remain a persistent barrier. Agentic AI systems must operate within existing organizational processes, legacy information systems, and regulatory constraints, yet their autonomous and adaptive behavior often conflicts with rigid workflows and hierarchical decision structures. Poor integration can result in duplicated effort, breakdowns in human–AI coordination, and reduced accountability, undermining both system performance and user trust. Addressing socio-economic and organizational barriers requires coordinated strategies that combine technical design with change management, institutional reform, and long-term investment in human capital, ensuring that agentic AI complements rather than destabilizes existing socio-technical systems.

\subsection{Future Directions}
This section delineates future directions for agentic AI by examining the coordinated evolution of technical capabilities, ethical and regulatory frameworks, socio-economic adaptation, and the long-term integration of agentic AI systems within socio-technical contexts.
\subsubsection{Technical Advancements}
Future technical advancements in agentic AI should prioritize system-level reliability, contextual intelligence, and governability rather than incremental gains in model performance. As agentic AI systems operate through long-horizon perception--planning--execution loops with limited human supervision, technical design choices directly shape societal risk, accountability, and trust. Three technical directions are particularly critical for supporting sustainable and responsible deployment.
\begin{description}
\item[Improved Robustness and Alignment.]
Robustness in agentic AI extends beyond tolerance to noisy inputs or distribution shifts. Autonomous agents continuously accumulate internal state, adapt plans, and modify behavior over time, allowing small errors or misaligned objectives to compound into systemic failures. Future work must therefore emphasize robustness mechanisms that constrain long-term behavior, including explicit safety boundaries, alignment-aware goal formulation, and monitoring of behavioral drift during extended operation. Rather than treating alignment as a static training objective, agentic AI requires dynamic alignment mechanisms that operate throughout the agent lifecycle and adapt to evolving environments and objectives.

\item[Multi-Modal and Context-Aware Systems.]
Agentic AI systems increasingly function in complex environments where isolated data streams are insufficient for informed decision-making. Multi-modal perception enables richer situational awareness, but technical progress must also address contextual understanding beyond immediate sensory inputs. Future systems should incorporate temporal, organizational, and institutional context into reasoning and planning processes, allowing agents to interpret actions within broader social and operational frameworks. Context-aware memory and selective information retention are essential to avoid brittle decision-making, loss of accountability, or uncontrolled accumulation of outdated knowledge.

\item[Secure and Governable Architectures.]
As autonomy increases, security and governability become foundational architectural requirements rather than auxiliary features. Agentic AI expands attack surfaces to include goal specifications, memory stores, tool interfaces, and inter-agent coordination mechanisms. Future architectures must therefore support continuous oversight, decision traceability, and controlled intervention without undermining operational effectiveness. Designing for governability requires built-in mechanisms for auditing, restricting autonomy, and integrating human oversight, ensuring that autonomous capabilities remain accountable to institutional, legal, and ethical constraints.
\end{description}

\subsubsection{Ethical and Regulatory Frameworks}
As agentic AI systems transition from decision-support tools to autonomous actors operating within social and institutional environments, ethical and regulatory frameworks must evolve accordingly. Traditional approaches to AI governance, which assume clearly bounded system behavior and direct human control, are increasingly insufficient for systems that plan, act, and adapt over extended time horizons. Future ethical and regulatory frameworks must therefore be tightly coupled with the technical realities of agentic autonomy.
\begin{description}
\item[Clear Accountability Structures.]
One of the most pressing challenges in governing agentic AI is the diffusion of responsibility across developers, deployers, users, and autonomous systems themselves. When agents independently generate plans and execute actions, attributing responsibility for unintended outcomes becomes non-trivial. Future frameworks must establish clear accountability structures that map autonomous decisions to human and institutional roles across the system lifecycle. This requires mechanisms for decision traceability, role differentiation, and responsibility assignment that remain effective even when agents adapt behavior dynamically. Accountability should be treated as a design requirement rather than a post hoc legal concern, ensuring that responsibility can be meaningfully exercised and enforced.

\item[Bias Mitigation and Inclusive Design.]
Bias in agentic AI systems arises not only from training data, but from how perception, reasoning, and planning components interact over time. Autonomous systems may amplify latent biases through feedback loops, selective memory, and goal optimization processes. Ethical frameworks must therefore promote inclusive design principles that address bias at multiple levels, including data collection, system objectives, and deployment contexts. Future approaches should emphasize continuous bias monitoring and adaptive mitigation rather than one-time fairness audits, ensuring that agentic systems remain responsive to evolving social norms and diverse stakeholder perspectives.

\item[Privacy-Preserving Technologies.]
Agentic AI systems rely heavily on continuous data acquisition, long-term memory, and contextual reasoning, intensifying privacy risks compared to conventional AI applications. Ethical and regulatory frameworks must support privacy-preserving technologies that limit unnecessary data exposure while enabling autonomous functionality. This includes mechanisms for data minimization, controlled retention, and user consent that operate throughout the agent’s lifecycle. Importantly, privacy should be integrated as a systemic property of agentic AI rather than an external compliance constraint, balancing personalization, autonomy, and individual rights.
\end{description}

\subsubsection{Socio-Economic Adaptation}
The long-term impact of agentic AI systems depends not only on technical feasibility or ethical soundness, but also on their integration into socio-economic and organizational structures. As agentic AI transitions from experimental deployments to sustained operation within enterprises, public institutions, and critical infrastructures, adaptation at the human, organizational, and economic levels becomes essential. Future research and policy efforts must therefore address how agentic autonomy reshapes collaboration, work practices, and value creation.
\begin{description}
\item[Human--AI Collaboration Models.]
Unlike traditional automation, agentic AI systems actively participate in decision-making, coordination, and goal formulation. This shift challenges established models of human--AI interaction that assume clear task boundaries and human dominance. Future collaboration models must redefine the division of labor between humans and autonomous agents, emphasizing complementarity rather than replacement. Effective human--AI collaboration requires mechanisms that allow humans to supervise, guide, and intervene in agent behavior while preserving the benefits of autonomy. Designing interaction paradigms that support shared situational awareness, mutual trust, and role clarity is critical to preventing over-reliance, disengagement, or loss of human agency.

\item[Workforce Reskilling and Management Training.]
The deployment of agentic AI alters skill requirements across organizational hierarchies. Beyond technical expertise, workers and managers must develop competencies in supervising autonomous systems, interpreting AI-driven decisions, and managing exceptions when systems fail or behave unexpectedly. Future socio-economic adaptation strategies should prioritize continuous reskilling and management training that align human capabilities with evolving agentic roles. Importantly, adaptation efforts must extend beyond frontline workers to include decision-makers responsible for governance, risk management, and accountability, ensuring that organizational leadership can effectively oversee autonomous operations.

\item[Cost--Benefit Analysis and Realistic ROI.]
Agentic AI systems often promise efficiency gains through autonomy and scalability, yet their real-world costs extend beyond development and deployment. Long-term operation requires investment in oversight, maintenance, governance, and human adaptation. Future evaluations must therefore adopt realistic cost--benefit analyses that account for organizational change, risk mitigation, and societal impact. Unrealistic expectations of rapid return on investment can lead to premature deployment or abandonment, undermining trust and sustainability. Transparent assessment frameworks are essential for aligning economic incentives with responsible adoption.
\end{description}

\subsubsection{Long-Term Societal Integration}
The long-term integration of agentic AI into society represents a qualitative shift in how intelligent systems participate in social, economic, and institutional processes. As agentic AI systems scale in autonomy and coordination, their impact extends beyond individual deployments toward the formation of persistent agentic networks embedded within societal infrastructures. Understanding this transition is essential for anticipating systemic effects that unfold over extended time horizons.
\begin{description}
\item[Agentic Networks and Ecosystems.]
Future agentic AI systems are unlikely to operate as isolated entities. Instead, they will increasingly form interconnected networks of agents that coordinate across organizations, sectors, and geographic boundaries. Such agentic ecosystems enable distributed decision-making, collective adaptation, and large-scale optimization, but they also introduce new forms of systemic risk. Interactions among autonomous agents can generate emergent behavior that is difficult to predict, attribute, or control at the level of individual components. Long-term societal integration, therefore, requires frameworks that address coordination, dependency, and resilience at the ecosystem level, rather than focusing solely on individual agent behavior.

\item[Cultural and Behavioral Shifts.]
As agentic AI becomes a routine participant in decision-making and coordination, it reshapes human expectations, social norms, and behavioral patterns. Repeated interaction with autonomous systems may influence how individuals perceive authority, responsibility, and expertise, potentially altering trust relationships and decision practices. Over time, societies may adapt by delegating increasing levels of judgment to agentic systems, raising concerns about over-reliance, erosion of human agency, and normalization of opaque decision-making. Addressing these cultural shifts requires sustained attention to transparency, education, and participatory governance, ensuring that societal adaptation remains intentional rather than implicit.

\item[Global Cooperation and Standards.]
The societal impact of agentic AI extends across national and regulatory boundaries, making unilateral governance approaches insufficient. Autonomous agents may operate within transnational infrastructures, global supply chains, or shared digital environments, where fragmented regulations create gaps in oversight and accountability. Long-term integration, therefore, depends on international cooperation and the development of shared standards that align technical design with common ethical and societal principles. While uniform regulation may be infeasible, interoperable governance frameworks can support consistency, trust, and coordination across jurisdictions, reducing the risk of regulatory arbitrage and uneven societal outcomes.
\end{description}

\section{Conclusions} \label{conclusion}
This paper analyzed core agentic AI components in relation to societal, ethical, economic, environmental, and governance considerations, highlighting the co-dependence between technical design decisions and the social contexts in which agentic systems operate. We introduced the MAD–BAD–SAD construct as an analytical lens that captured this interdependence by clarifying how the MAD, BAD, and SAD dimensions arose from the interaction between agentic AI capabilities and real-world socio-technical conditions. We discussed ethical considerations, implications, and challenges arising from current technical developments in agentic AI through the proposed MAD, BAD, and SAD dimensions. We assessed these developments across a range of application domains in which agentic AI is actively emerging, including healthcare, education, industry, smart and sustainable cities, social services, communications and networking, and earth observation and satellite communications for resilience and long-term adaptation. We further examined agentic AI as an integrated socio-technical system whose behavior and impact were co-produced by algorithms, data, organizational practices, regulatory frameworks, and social norms. Finally, we outlined several future research directions aimed at addressing these challenges and advancing socio-technical approaches to agentic AI that support long-term sustainability and responsible integration into society.


\bibliographystyle{ACM-Reference-Format}
\bibliography{ref}

\end{document}